\documentclass[twocolumn,letterpaper]{IEEEAerospaceCLS}  

\usepackage[]{graphicx}    
\usepackage{cite}
\usepackage{caption}
\usepackage{subcaption}
\usepackage{subfig}
\usepackage{float}
\usepackage{dblfloatfix}
\newcommand{\ignore}[1]{}  

\begin{document}
\title{The Optical Design of the Carbon Investigation (Carbon-I) Imaging Spectrometer}

\author{%
Christine L. Bradley, Rami W. Wehbe, Matthew Smith, Sharmila Padmanabhan,\\ Valerie Scott, David R. Thompson, Daniel W. Wilson, Pantazis Mouroulis, Robert O. Green\\ 
Jet Propulsion Laboratory, California Institute of Technology\\
4800 Oak Grove Drive\\
Pasadena, CA 91109\\
christine.l.bradley@jpl.nasa.gov
\and 
Christian Frankenberg\\
Division of Geological and Planetary Sciences, California Institute of Technology\\
4800 Oak Grove Dr.\\
Pasadena, CA 91109
\thanks{\footnotesize   979-8-3503-5597-0/25/$\$31.00$ \copyright2025 IEEE}              
}

\maketitle

\thispagestyle{plain}
\pagestyle{plain}

\maketitle

\thispagestyle{plain}
\pagestyle{plain}

\begin{abstract}
The proposed Carbon Investigation (Carbon-I) Imaging Spectrometer is designed to measure variations of greenhouse gases in the Earth’s atmosphere. The instrument will survey the Earth from its own spacecraft at an altitude of approximately 610 km. It will use a coarse ground sampling distance (GSD) of $<$400 m in global mode for land and coastal monitoring and a finer 35 m GSD in target mode to sample key regions. The identification and quantification of greenhouse gases require continuous spectral sampling over the 2040-2380 nm wavelength range with $<$1 nm spectral sampling. The proposed design builds upon Jet Propulsion Laboratory’s (JPL) experience of spaceflight Dyson imaging spectrometers to achieve spectral sampling of 0.7 nm per pixel. This paper presents the proposed Carbon-I optical design comprised of a freeform three-mirror anastigmat telescope that couples to a F/2.2, highly uniform Dyson-inspired imaging spectrometer. The high uniformity and throughput enables Carbon-I to measure Earth’s greenhouse gas concentrations with unprecedented precision and spatial sampling. 
\end{abstract} 

\begin{table*}
\renewcommand{\arraystretch}{1.3}
    \centering
\caption{\bf Optical Design Properties and Requirements comparison between Dyson Imaging Spectrometers Carbon-I, SBG VSWIR, and EMIT.}
\label{tab:opticalDesign}
    \begin{tabular}{|p{0.22\textwidth}|p{0.225\textwidth}|p{0.225\textwidth}|p{0.225\textwidth}|} \hline 
         \bfseries Performance Parameter&  \bfseries Carbon-I&  \bfseries SBG VSWIR&  \bfseries EMIT\\ \hline \hline
         Configuration&  Single telescope and Dyson inspired spectrometer&  Two identical telescope and Dyson inspired spectrometer assemblies&  Single telescope and Dyson spectrometer\\ \hline 
         Status&  Proposal phase&  Preliminary design phase&  Operating since July 2022\\ \hline 
         FOV&  $\ge$ 8.9 deg&  $\ge$ 17 deg&  11 deg\\ \hline 
         Ground Sampling Distance&  $\le$ 50 m&  30-35 m&  60-65 m\\ \hline 
         F/$\#$ & 2.2& 1.8&1.8\\ \hline 
         Slit Length&  $\ge$ 54 mm&  $\ge$ 54 mm&37.2 mm\\ \hline 
         Slit Width& 36 $\mathrm{\mu}$m (2x pixel)& 36 $\mathrm{\mu}$m (2x pixel)&30 $\mathrm{\mu}$m\\ \hline
         Pixel Size& 18 $\mathrm{\mu}$m& 18 $\mathrm{\mu}$m in spatial dim. co-add 2 pixels in spectral&30 $\mathrm{\mu}$m\\ \hline 
         Detector Array Size& 3072 x 512& 3072 x 512&1280 x 480\\ \hline 
         Spectral Range& 2.04 - 2.368 $\mathrm{\mu}$m& 0.38 - 2.5 $\mathrm{\mu}$m&0.38 - 2.5 $\mathrm{\mu}$m\\ \hline 
         Spectral Sampling& $\le$ 1 nm& $\le$ 10 nm per co-added pixel&7.4 nm\\ \hline 
         Telescope Temperature& 275 - 305 K& 280 - 300 K&245 - 300 K\\ \hline 
         Spectrometer Temperature& 238 - 242 K& 238 - 242 K&238 - 242 K\\ \hline 
         FPA Temperature&  150 - 160 K& 150 - 160 K&150 - 160 K\\ \hline
    \end{tabular}
\end{table*}

\tableofcontents

\section{Introduction}
The Carbon Investigation (Carbon-I) imaging spectrometer is proposed to monitor greenhouse gases (GHG), particularly methane (CH$_4$) and carbon dioxide (CO$_2$), to understand contributions from natural (e.g. tropical wetlands) and anthropogenic sources such as waste management, agriculture, and fossil fuel production. Due to the recent acceleration of methane in the atmosphere, there is an increased need to understand and distinguish natural climate feedbacks and anthropogenic contributions. While recent space-based GHG measurements have improved the understanding of the gas exchange rates between the Earth’s surface and atmosphere \cite{Frankenberg2005,Crisp2004,Bovensmann1999,Butz2011,Kuze2016,Lorente2021,Jacob2022}, these measurements are obtained with a spatial sampling too coarse to identify the GHG fluxes in the humid tropics, where shallow cumulus clouds are prevalent \cite{frankenberg2024data}. The tropics exhibit the largest carbon fluxes and uncertainties but the lowest data coverage from current satellites. A new spectrometer approach is thus required to fill the data gap in the tropics.

The Carbon-I imaging spectrometer enables identification of sources by providing high spectral sampling and resolution while imaging at high spatial resolutions. Carbon-I supports multi-species measurements of CH$_4$, CO$_2$, and CO by sampling at 0.7\,nm over the 2040-2380\,nm spectral range. Carbon-I operates in two modes, global mode and target mode, that enable precise localization, quantification, and attribution of GHG fluxes at local and global scales. The instrument will map all land surfaces monthly in global mode with a ground sampling distance (GSD) of $<$400\,m and a 100\,km swath width. A finer GSD of 35 m in target mode will map specific regions of interest over a 100\,km x 100\,km area. The Carbon-I measurements will be used to constrain global GHG fluxes and bridge the scales from local to global. These data will support the three key mission objectives:

\begin{enumerate}

  \item Identify CH$_4$ and CO$_2$ emission hotspots across the globe. 
  \item Quantify global monthly CH$_4$, CO$_2$, and CO fluxes at an unprecedented 12 - 100\,km spatial resolution.
  \item Attribute and quantify the processes driving both natural and anthropogenic emissions.
\end{enumerate}

\begin{table*}[!b]
\renewcommand{\arraystretch}{1.3}
    \centering
\caption{\bf Optical Performance Requirements comparison between Dyson Imaging Spectrometers Carbon-I, SBG VSWIR, and EMIT. The Along-track Response Function (ARF), Spectral Response Function (SRF), Cross-track Response Function (CRF) full width at half-maximum (FWHM), Smile, and Keystone requirements are represented in pixel units and converted to spatial dimension units for comparison across instruments. }
\label{tab:opticalRequirements}
    \begin{tabular}{|p{0.11\textwidth}  p{0.11\textwidth}|p{0.22\textwidth}|p{0.22\textwidth}|p{0.22\textwidth}|} \hline 
         \multicolumn{2}{|c|}{\bfseries Performance Parameter}& \bfseries Carbon-I&  \bfseries SBG VSWIR&  \bfseries EMIT\\ \hline \hline
         ARF FWHM&  pixel units&$\le$ 3.0 pixels&  $\le$ 2.8 pixels&  $\le$ 1.7 pixels\\ 
         & spatial units& $\le$ 54\,$\mathrm{\mu}$m for 18\,$\mathrm{\mu}$m pixel&  $\le$ 50.4\,$\mathrm{\mu}$m for 18 $\mathrm{\mu}$m pixel&  $\le$ 51 $\mathrm{\mu}$m for 30 $\mathrm{\mu}$m pixel\\ \hline
         SRF FWHM& spectral units& $\le$ 2.5 nm& $\le$ 18 nm&$\le$ 12.58 nm\\
         & pixel units& $\le$ 2.5 pixels for 1 nm/pix& $\le$ 1.8 pixels for 5 nm/pix&$\le$ 1.7 pixels for 7.5 nm/pix\\
         & spatial units& $\le$ 45 $\mathrm{\mu}$m for 18 $\mathrm{\mu}$m pixel& $\le$ 64.8 $\mathrm{\mu}$m for 36 $\mathrm{\mu}$m pixel&$\le$ 51 $\mathrm{\mu}$m for 30 $\mathrm{\mu}$m pixel\\\hline
         CRF FWHM& pixel units& $\le$ 2.5 pixels& $\le$ 2.8 pixels&$\le$ 1.7 pixels\\
         & spatial units& $\le$ 45 $\mathrm{\mu}$m for 18 $\mathrm{\mu}$m pixel& $\le$ 50.4 $\mathrm{\mu}$m for 18 $\mathrm{\mu}$m pixel&$\le$ 51 $\mathrm{\mu}$m for 30 $\mathrm{\mu}$m pixel\\\hline
         Smile& pixel units& $\le$ 15$\%$ of pixels& $\le$ 5$\%$ of co-added pixels&$\le$ 10$\%$ of pixels\\
         & spatial units& $\le$ 2.7$\mathrm{\mu}$m for 18 $\mathrm{\mu}$m pixel& $\le$ 1.8$\mathrm{\mu}$m for 36$\mathrm{\mu}$m pixel&$\le$ 3.0$\mathrm{\mu}$m for 30 $\mathrm{\mu}$m pixel\\\hline
         Keystone& pixel units& $\le$ 15$\%$ of pixels& $\le$ 10$\%$ of pixels&$\le$ 10$\%$ of pixels\\
         & spatial units& $\le$ 2.7$\mathrm{\mu}$m for 18 $\mathrm{\mu}$m pixel& $\le$ 1.8$\mathrm{\mu}$m for 18$\mathrm{\mu}$m pixel&$\le$ 3.0$\mathrm{\mu}$m for 30 $\mathrm{\mu}$m pixel\\\hline 
    \end{tabular}
\end{table*}

This paper details the Carbon-I optical design that achieves the high spatial and spectral sampling required to meet the mission objectives by demonstrating engineering solutions to the design requirements. The Carbon-I optical design utilizes lessons learned from previously developed imaging spectrometers at Jet Propulsion Laboratory (JPL) such as the NASA Earth Surface Mineral Dust Source Investigation (EMIT) instrument \cite{2018AGUFM.A24D..01G,Green2020,Bradley2020,THOMPSON2024113986} as well as the Orbiting Carbon Observatories (OCO, OCO-2, and OCO-3) \cite{MILLER2005876,Day5723741,amt-10-59-2017,Lee7835630,Keller9928530}. EMIT identifies mineral compositions of desert regions on Earth with a GSD of 60 m from the vantage point of the International Space Station with a Dyson imaging spectrometer that operates over 380-2500 nm at a spectral sampling of 7.4 nm. Carbon-I achieves 10x finer spectral sampling and 2x finer ground sampling distance with a freeform three-mirror anastigmat telescope that feeds light into an optically fast F/2.2 Dyson-inspired imaging spectrometer. JPL's advancements in imaging spectrometer development enables finer spectral and spatial resolution capability to achieve the requirements of the Carbon-I mission. 

\section{Optical Performance Requirements and Design Overview}

\begin{figure*}
\centering
\includegraphics[width=\textwidth]{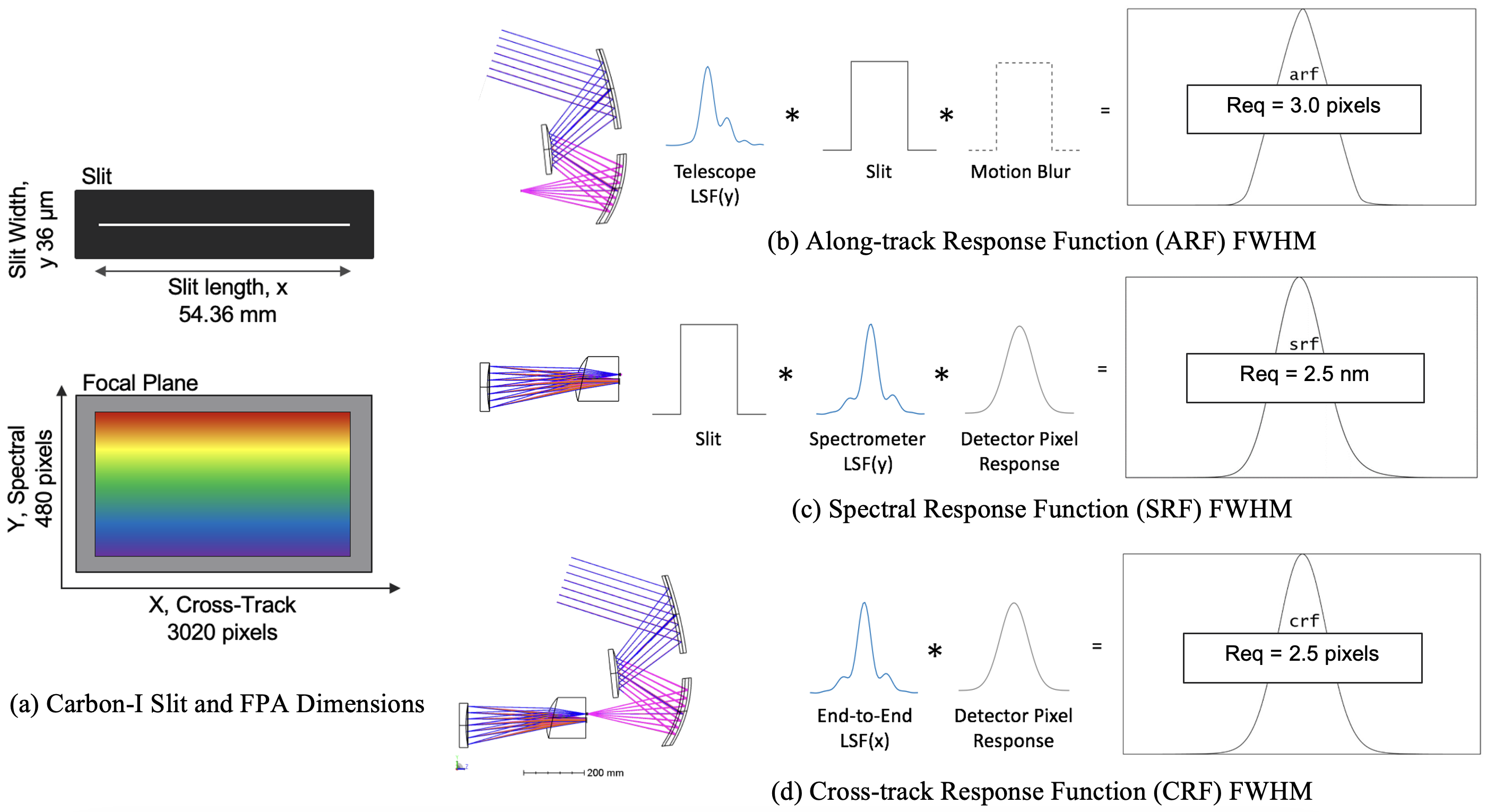}
\caption{\bf{(a) Configuration of Carbon-I slit and FPA where the slit length follows the FPA spatial dimension and the slit width corresponds to the dispersion direction on the FPA. Optical resolution calculations and requirements are (b) the Along-track Response Function (ARF), (c) Spectral Response Function (SRF), and (d) the Cross-track Response Function (CRF). These calculations only apply in the incoherent approximation.}}
\label{fig:requirement}
\end{figure*}

In addition to EMIT, Carbon-I builds upon experiences from the Earth Science Decadal Survey imaging spectrometer in development called Surface Biology Geology (SBG) Visible to Shortwave Infrared (VSWIR) Wide Swath Imaging Spectrometer \cite{Green9843676,Bradley10.1117/12.2692105}. Comparisons of the Carbon-I specifications to the EMIT instrument that has been operating since July 2022 and the SBG VSWIR instrument that is currently in the Preliminary Design phase are shown in Table \ref{tab:opticalDesign}. EMIT and imaging spectrometers before SBG VSWIR used a smaller focal plane array (FPA) size that forced trade-offs between swath width and ground sampling distances that resulted a larger GSD of 60 m for the EMIT instrument. Carbon-I and SBG VSWIR use a new Teledyne Chroma-D detector that has the larger 3072 x 512 size that provides similar swath width coverage as EMIT but at the finer 30 m GSD. Carbon-I is comprised of a single telescope and spectrometer assembly to provide approximately half the swath of SBG VSWIR. The Carbon-I spectrometer requires a long slit to map to the long dimension of the FPA, which pushes the spectrometer design to be larger than previous designs. The primary difference between Carbon-I and SBG VSWIR is the fine spectral sampling over a narrower spectral range required for CH$_4$ and CO$_2$ studies. Carbon-I also achieves higher optical throughput to help suppress stray light impacts. 

\begin{figure*}[!t]
    \begin{subfigure}[b]{0.5\textwidth}
        \centering
        \includegraphics[height=2.8in]{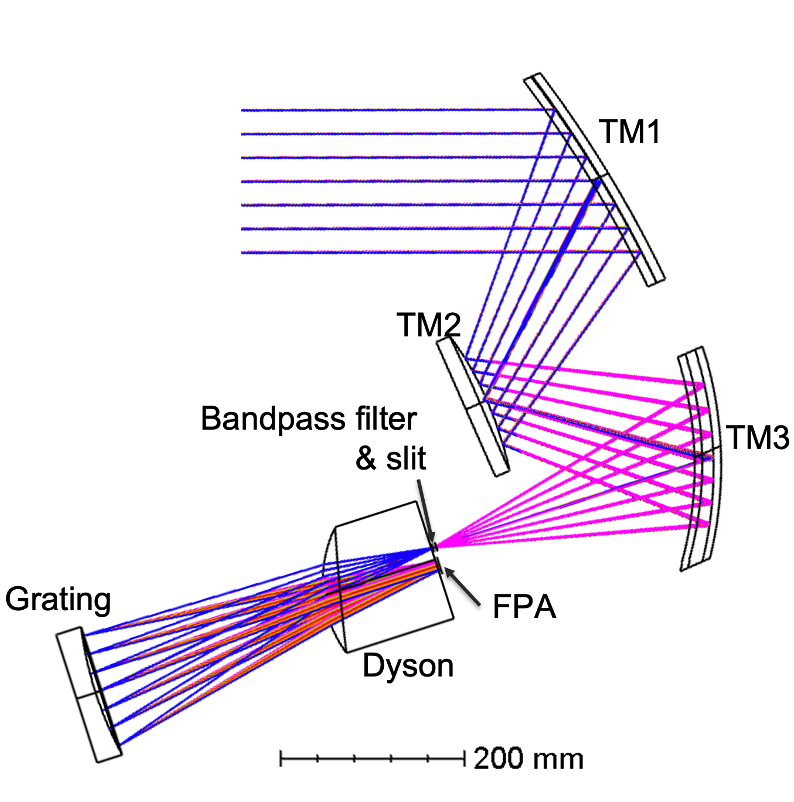}
        \caption{Carbon-I Optical Ray Trace}
    \end{subfigure}
    \begin{subfigure}[b]{0.5\textwidth}
        \centering
        \includegraphics[height=2.8in]{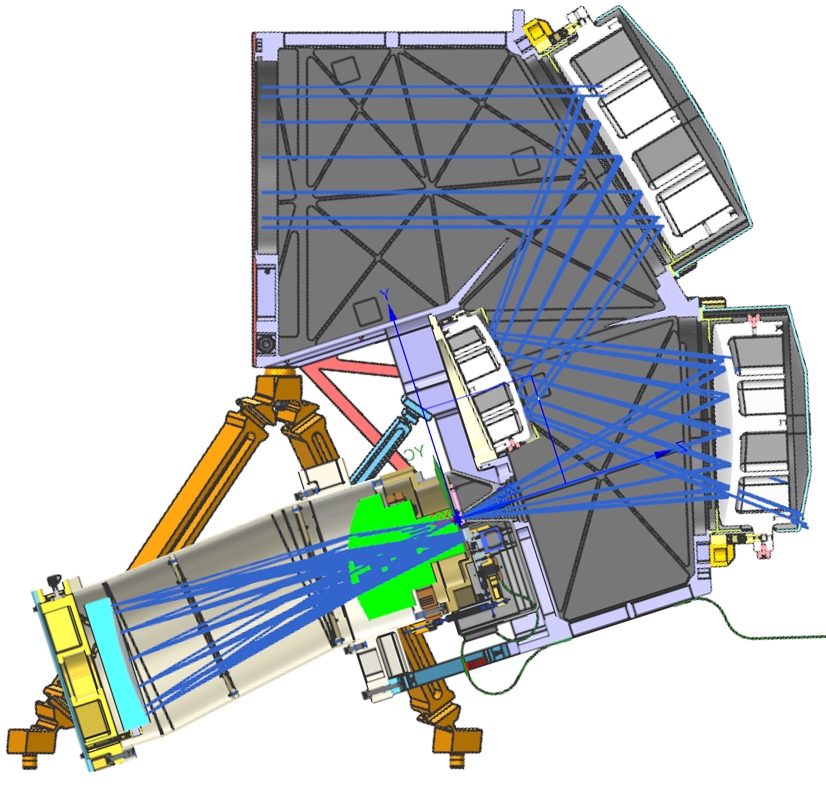}
        \caption{Carbon-I Proposal Structural Design}
    \end{subfigure}
    \caption{\bf{Carbon-I (a) optical ray trace and (b) the proposed structural design.}}
    \label{fig:raytrace}
\end{figure*}

\begin{figure*}[!b]
    \begin{subfigure}[b]{0.5\textwidth}
        \centering
        \includegraphics[height=2.6in]{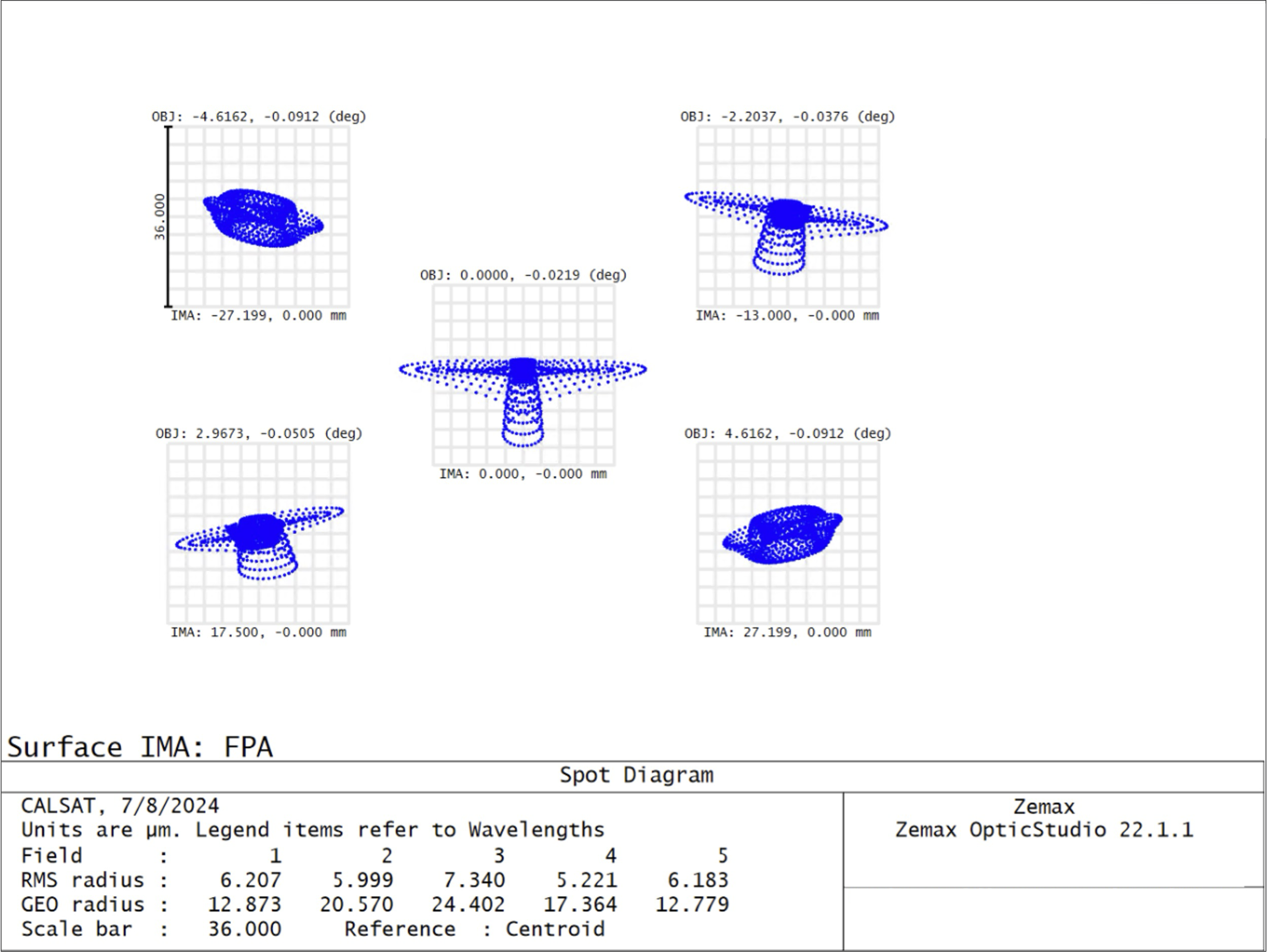}
        \caption{Spot diagrams for three-mirror anastigmat telescope with even aspheric mirrors.}
    \end{subfigure}
    \begin{subfigure}[b]{0.5\textwidth}
        \centering
        \includegraphics[height=2.6in]{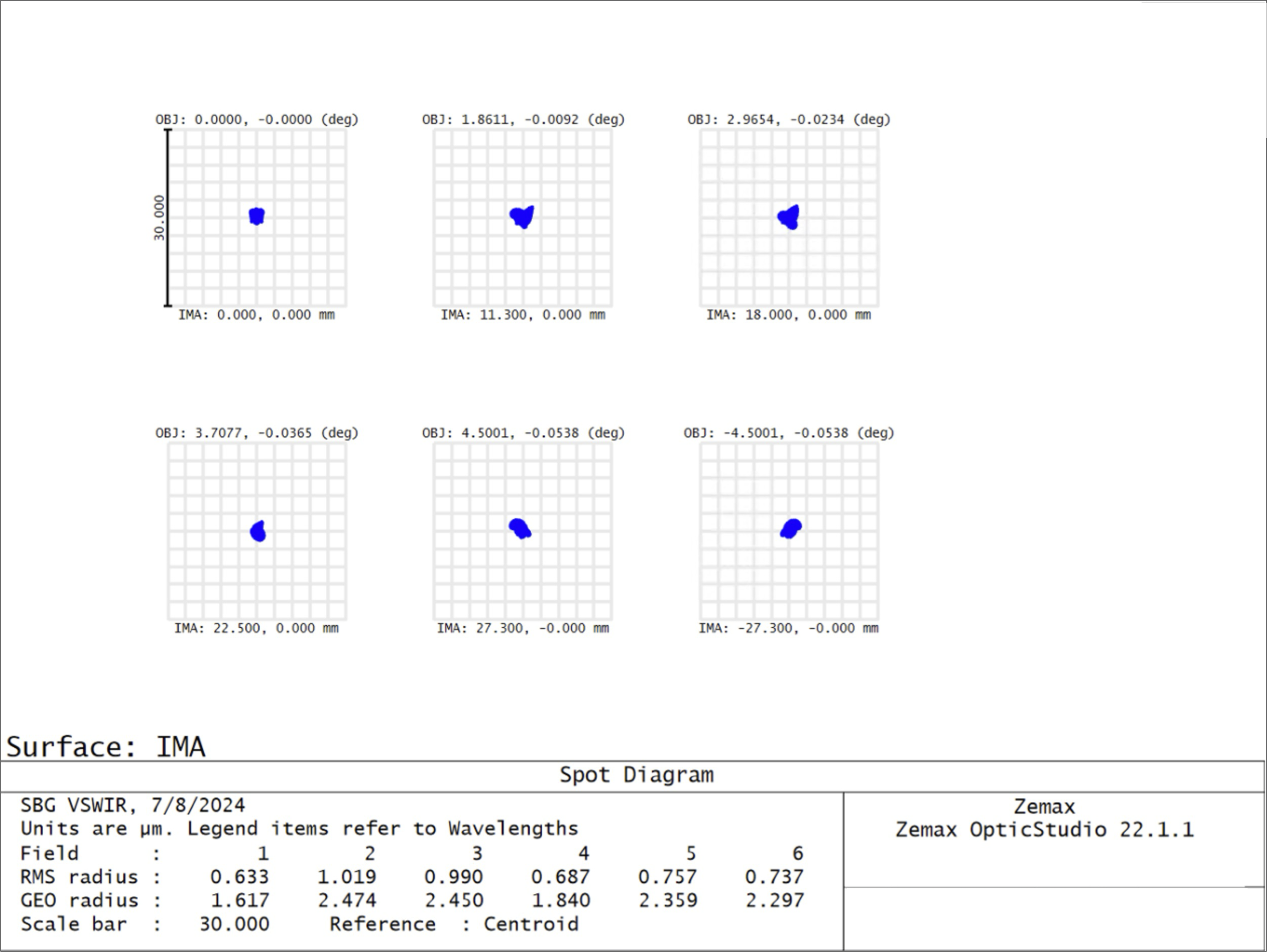}
        \caption{Spot diagrams for SBG VSWIR three-mirror anastigmat telescope with freeform mirrors.}
    \end{subfigure}
    \caption{\bf{Telescope design trade options for (a) an even aspheric three-mirror anastigmat telescope and (b) the SBG VSWIR Freeform three-mirror anastigmat telescope. The spots presented for the two options are on the same scale where each box is 2-pixels by 2-pixels (36\,$\mathrm{\mu}$m by 36\,$\mathrm{\mu}$m). The two telescope options are similar in size but the freeform telescope compensates for the aberrations across the field of view to deliver diffraction limited spots.}}
    \label{fig:telescopespots}
\end{figure*}

The optical design performance metrics are detailed in Figure \ref{fig:requirement} and further presented by Mouroulis and Green \cite{Mouroulis10.1117/1.OE.57.4.040901}. The spatial direction corresponds to the long dimension of the array. There are three resolution requirements that define three subsystems of the optical design: the telescope, the spectrometer, and the full end-to-end system. The Along-track Response Function (ARF) defines the telescope only performance in the system. The spectrometer only performance metric is defined as the Spectral Response Function (SRF). Finally, the Cross-track Response Function (CRF) defines the full end-to-end performance from the telescope to the spectrometer. Note that the ARF and SRF calculations are partly defined by the line spread function (LSF) in the y-direction of the telescope and spectrometer, respectively. The CRF is partly defined by the LSF in the x-direction, perpendicular to ARF and SRF, of the full end-to-end system. The calculations of the response functions apply in the incoherent approximation to simplify computations. Previous imaging spectrometer experience have shown this approximation agrees well with as-built hardware. 

Typically margins are calculated by normalizing the difference between the current best estimate (CBE) from the requirement by the requirement value. However, the response function margins are calculated differently to exclude the contribution of the slit width or the detector pixel response width. The response functions are defined by convolutions with the slit width or detector pixel width in the incoherent approximation such that the response functions can never be less than either the slit width or detector pixel width. The margins calculations are

\begin{equation}
    \textnormal{Margin}_{\textnormal{SRF}} = \frac{\left(\textnormal{Requirement} - \textnormal{CBE}\right)}{\left(\textnormal{Requirement} - 2\times \Delta\lambda\right)},
\end{equation}

\begin{equation}
    \textnormal{Margin}_{\textnormal{ARF}} = \frac{\left(\textnormal{Requirement} - \textnormal{CBE}\right)}{\left(\textnormal{Requirement} - 2\times \Delta x\right)},
\end{equation}

\begin{equation}
    \textnormal{Margin}_{\textnormal{CRF}} = \frac{\left(\textnormal{Requirement} - \textnormal{CBE}\right)}{\left(\textnormal{Requirement} - \Delta x\right)},
\end{equation}

where $\Delta\lambda$ is the spectrometer spectral sampling of 0.7\,nm and $\Delta x$ is the pixel size of 18\,$\mathrm{\mu}$m. Twice the spectral sampling and twice the pixel size are excluded from SRF and ARF, respectively, because the slit width is twice the size of a pixel. 

The instrument has two additional optical performance requirements that define the centroid uniformity across the FPA: smile and keystone. The smile requirement defines that deviation of centroids for a given wavelength across the spatial dimension of the focal plane cannot exceed more than 15$\%$ of a pixel. The keystone performance metric defines the deviation of centroids in the orthogonal direction where the spectra for a given field point cannot deviate more than 15$\%$ of a pixel. 

The Carbon-I optical design balances the five optical performance metrics while meeting first order optical design requirements. Table \ref{tab:opticalRequirements} lists the optical performance requirements for Carbon-I, SBG VSWIR, and EMIT. To compare across instruments, the requirements are converted from pixel units that can vary depending on the FPA pixel size or from the differing spectral sampling to a common spatial unit. It is shown that Carbon-I has a challenging SRF and CRF FWHM requirements of 45 $\mathrm{\mu}$m compared to the SBG VSWIR and EMIT SRF FWHM and CRF FWHM of $\ge$51 $\mathrm{\mu}$m. The Carbon-I SRF FWHM is less sensitive than the Carbon-I CRF because we want to oversample the spectral response function with at least two detector pixels. The centroid uniformity requirements are similar to what was achieved for EMIT. 

We thus optimize the optical design given the SRF and CRF requirements to provide a point design with substantial margin to allocate to as-built and in-flight stability errors. Both the telescope and spectrometer contribute to the CRF performance. The telescope design is optimized separately from the spectrometer to provide small spot sizes that are uniform across the full field of view. While the SRF requirement is key for this mission, we show that it is less sensitive to errors compared to the CRF requirement. The initial Carbon-I point design thus slightly favors the CRF in the x-direction in order to provide healthy margin for all optical performance requirements. The resulting proposed optical configuration and structure design shown in Figure \ref{fig:raytrace} utilizes a freeform telescope coupled with a CRF optimized Dyson-inspired imaging spectrometer.

\begin{figure*}[!t]
    \begin{subfigure}[b]{0.5\textwidth}
        \centering
        \includegraphics[height=3.3in]{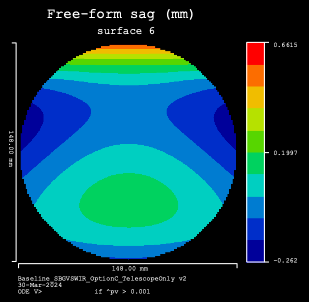}
        \caption{Freeform sag departure from the best fit sphere for the secondary telescope mirror.}
    \end{subfigure}
    \begin{subfigure}[b]{0.5\textwidth}
        \centering
        \includegraphics[height=3.3in]{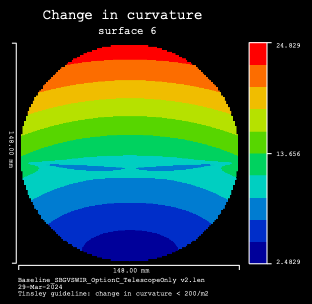}
        \caption{The derivative of the secondary telescope mirror curvature is 25 m$^{-2}$, well under the vendor's manufacturing limit.}
    \end{subfigure}
    \caption{\bf{Guidance for manufacturing feasibility of freeform mirrors is that the derivative of the curvature needs to be less than 200 m$^{-2}$. The secondary mirror in the SBG VSWIR telescope shows the largest change in curvature for all three mirrors. The SBG VSWIR telescope mirrors are well within the the vendor's manufacturing ability.}}
    \label{fig:curvatures}
\end{figure*}

\section{Telescope Design}
Early in formulation, a telescope trade was performed between an even aspheric TMA telescope and a freeform TMA telescope designed for the SBG VSWIR instrument. The original even aspheric TMA telescope has similar focal length as the SBG VSWIR freeform telescope where both meet the altitude, swath width, and ground sampling distance requirements for the Carbon-I mission. A comparison of the spot sizes for the two telescopes is shown in Figure \ref{fig:telescopespots} where (a) is the even aspheric telescope and (b) is the SBG VSWIR freeform telescope. The SBG VSWIR freeform telescope similar in size to the aspheric telescope but delivers uniform and small spots across the instrument field of view. This results in better CRF FWHM design performance and a telescope that is less sensitive to fabrication, alignment, and in-flight errors. The SBG VSWIR telescope has the advantage that it was designed at F/1.8 to overfill the F/2.2 Carbon-I spectrometer. The smaller cone of rays allows for efficient placement of internal telescope vanes for stray light control that also result in an efficient fore-baffle.

\begin{figure*}[!t]
    \begin{subfigure}[b]{0.5\textwidth}
        \centering
        \includegraphics[height=2.6in]{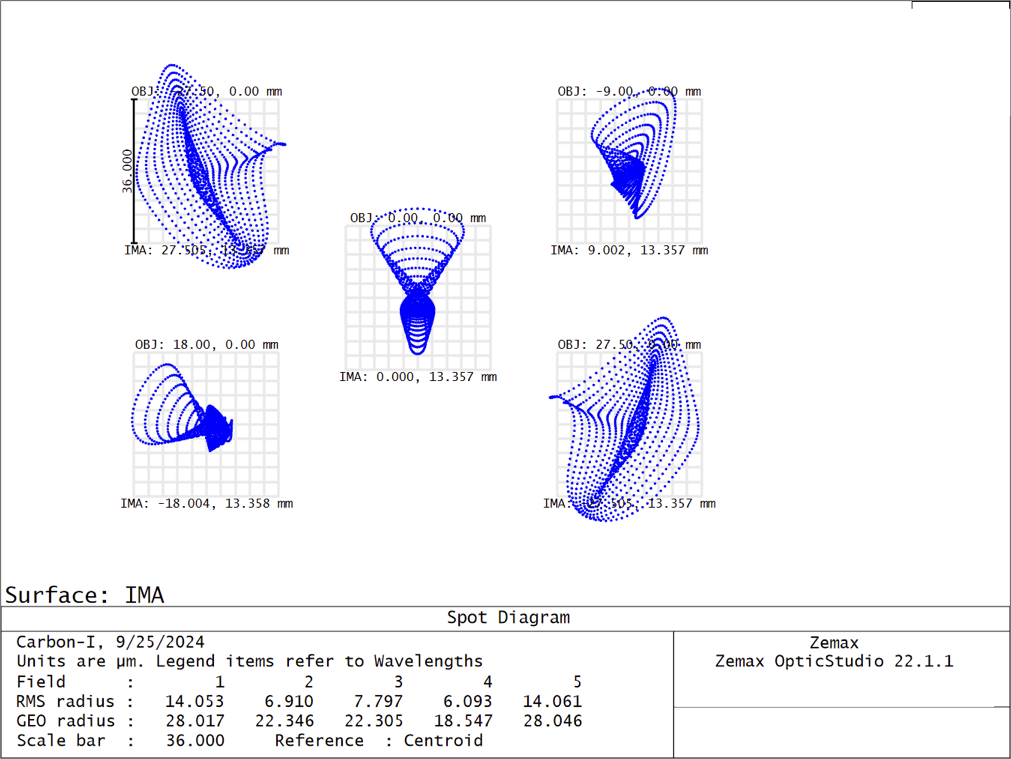}
        \caption{Spectrometer spot diagrams at 2040 nm.}
    \end{subfigure}
    \begin{subfigure}[b]{0.5\textwidth}
        \centering
        \includegraphics[height=2.6in]{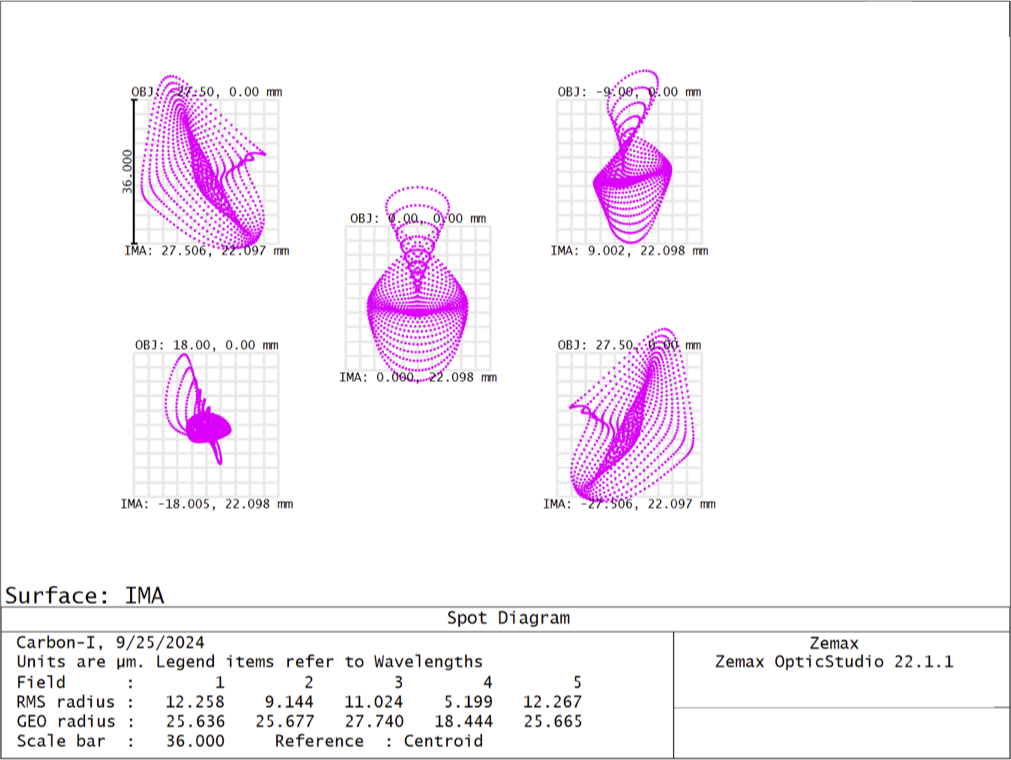}
        \caption{Spectrometer spot diagrams at 2380 nm.}
    \end{subfigure}
    \caption{\bf{Carbon-I spectrometer spot diagrams at (a) 2040 nm and (b)2380 nm. Each box is 2-pixels by 2-pixels (36 $\mathrm{\mu}$m by 36 $\mathrm{\mu}$m). Spots were optimized to be thinner in the x-direction to meet the CRF FWHM requirement.}}
    \label{fig:spectrometerSpots}
\end{figure*}

In addition to meeting first order design and optical performance requirements, the main advantage of the SBG VSWIR Telescope is that it is has passed the Optical and Optomehechanical Preliminary Design Reviews. The SBG VSWIR telescope will largely be build-to-print to allow for cost savings in design development and alignment planning activities. 

Both aspheric and freeform surfaces are manufactured and polished using the same process. Manufacturing feasibility guidance from the aluminum mirror fabrication vendor, Coherent, shows that mirrors with a derivative of curvature less than 200\,m$^{-2}$ can be fabricated with their current tooling. The max derivative of curvature for SBG VSWIR is observed for the secondary mirror at 25\,m$^{-2}$. Figure \ref{fig:curvatures} shows the maximum departure of the freeform surface from the best fit sphere and its corresponding derivative of curvature of the secondary telescope mirror. The SBG VSWIR telescope mirrors are well within the vendor's standard processing. 

The telescope is baselined to be all-aluminum with aluminum structure and aluminum mirrors. While the telescope structure and mirrors with lightweighting will remain the same, there are some small changes the Carbon-I team can make to improve upon stray light control as mentioned above and simplify the fabrication of the Carbon-I telescope mirrors. Since Carbon-I operates at a much longer wavelength than SBG VSWIR, the mirrors are protected gold coated instead of the protected silver coated mirrors required for SBG VSWIR. Carbon-I also has more relaxed mirror roughness requirements (40 Angstroms rms vs. 20 Angstroms rms), since it operates down to 2000\,nm where SBG VSWIR works down to 380\,nm.

\section{Spectrometer Design}

Jet Propulsion Laboratory has experience developing Dyson imaging spectrometers such as the Compact Wide Swath Imaging Spectrometer (CWIS) \cite{VanGorp10.1117/12.2062886,VanGorp10.1117/12.2239080}, Snow and Water Imaging Spectrometer (SWIS) \cite{Bender10.1117/1.JRS.12.044001}, EMIT \cite{Green2020,Bradley2020}, Mapping Imaging Spectrometer for Europa (MISE) \cite{Bender10.1117/12.2530464}, Carbon Plume Mapper (CPM) \cite{Zandbergen10.1117/12.2678614,Shaw10.1117/12.2678750}, and SBG VSWIR \cite{Green9843676,Bradley10.1117/12.2692105}. Previous Dyson imaging spectrometers exhibit large spectral ranges typically in the visible to shortwave infrared with spectral samplings on the order of 10\,nm per pixel. While EMIT can pinpoint methane sources, its spectral sampling is unable to effectively resolve the fine absorption features of greenhouse gases. The Carbon-I design will improve upon GHG detection by using a spectral sampling of\,0.7\,nm, achieving a resolving power ($\lambda/\Delta\lambda$) larger than 1000 and greatly increasing its sensitivity to GHG variations and providing separation of surface features and spectrally sharp atmospheric absorption lines. 

While a traditional Dyson imaging spectrometer is comprised of a Dyson lens with a convex spherical surface that is concentric with a concave spherical grating \cite{Dyson:59,Mouroulis10.1117/12.524626}, the Carbon-I spectrometer utilizes an even aspheric convex surface on the Dyson lens made of fused silica. The optical ray trace in Figure \ref{fig:raytrace}a shows the Dyson lens also is an off-axis part; however, its parent vertex is still on-axis with the grating to use the same alignment methodology as previous imaging spectrometers developed by JPL. When coupled with the grating groove line spacing, the slit and FPA decenter positions allow the spectrometer design to achieve the first order optical design property of 0.7\,nm spectral sampling. 

The Carbon-I grating is simplified from previous imaging spectrometers that operate over a large spectral range. Gratings from EMIT, CPM, and SBG VSWIR require shaped groove gratings to optimize for light at the longer wavelength end where there is less light in the solar spectrum. This results in low efficiency at the visible wavelength end and high efficiency at the longer wavelengths of the spectrum. The Carbon-I spectral range is narrow compared to previous imaging spectrometers and enables a triangular blazed grating optimized for the center wavelength of the spectral range. This provides a grating efficiency of $>$0.84 over the entire Carbon-I spectrum, resulting in much larger optical throughput than other broadband imaging spectrometers. This narrow spectral range also enables the efficient use of a bandpass filter to mitigate stray light due to wavelengths outside of the Carbon-I waveband. 

As mentioned in the previous section, the SRF and CRF FWHM requirements are tighter than previous instruments. Since the CRF depends on both the telescope and spectrometer performance and therefore more susceptible to error, the spectrometer was optimized to slightly favor the CRF requirement to ensure margins for all requirements after fabrication, alignment, and in-flight errors are accounted. The result can be seen in the spot diagrams presented in Figure \ref{fig:spectrometerSpots} as generally the spots are elongated in the y-direction. Each box in the spot diagrams corresponds to a 2-pixel by 2-pixel size. The RMS radius of the spots vary by 6-15 $\mathrm{\mu}$m across the spectral range but are narrower in the x (spatial) direction. This optimized configuration allows for healthy margins across all requirements. 

\section{End-to-End Performance}

\begin{figure*}[!t]
\centering
\includegraphics[width=\textwidth]{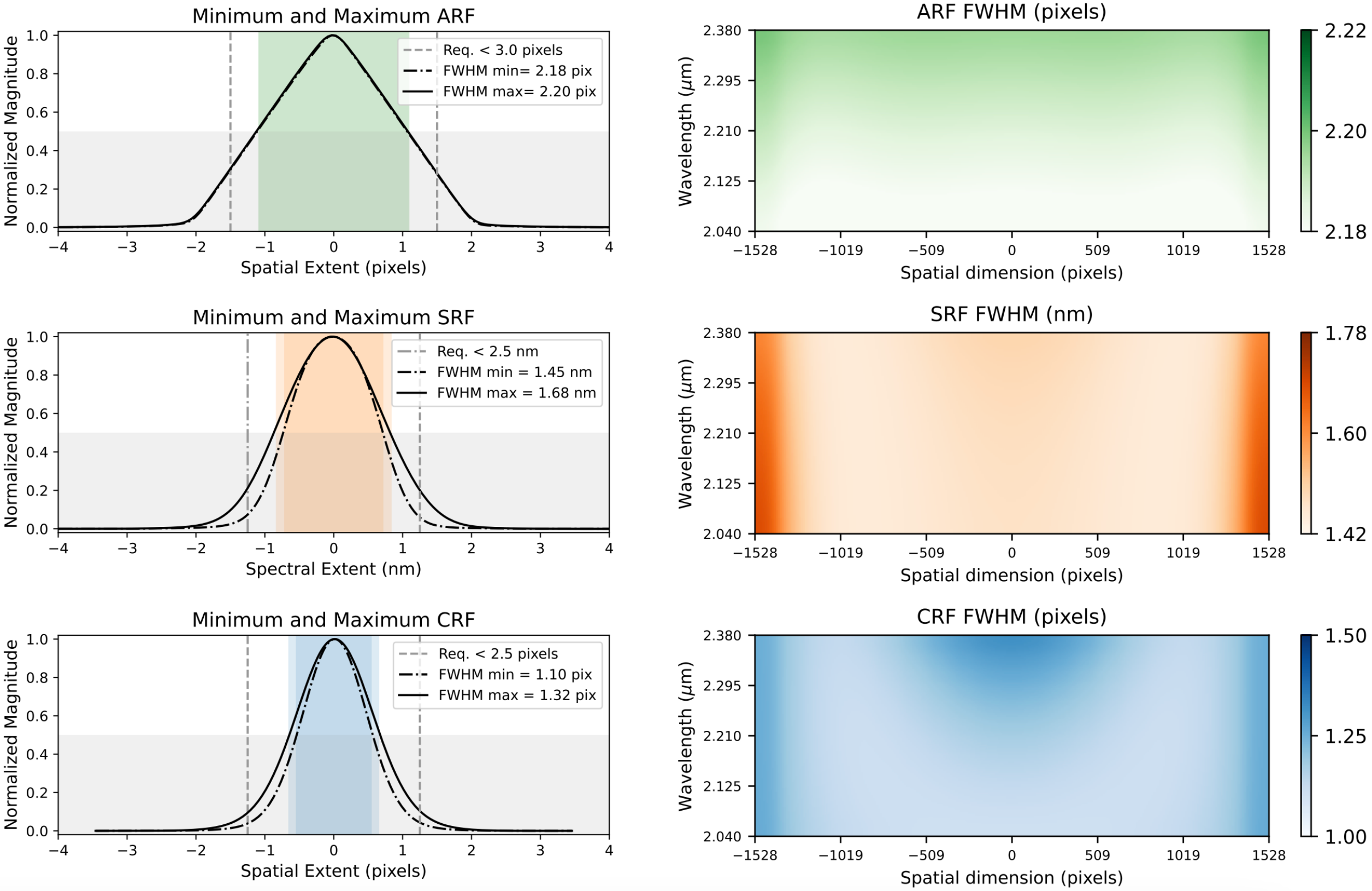}
\caption{\bf{The Carbon-I Design Along-track Response Function (top, green), Spectral Response Function (middle, orange), and Cross-Track Response Function (bottom, blue) are represented over fields and wavelengths.}}
\label{fig:pointRFs}
\end{figure*}

Both the spectrometer and telescope are designed and evaluated individually. Once designed independently, they are brought together to evaluate the full system performance through the response functions, smile, and keystone calculations. The resulting ARF, SRF, and CRF values of the resulting end-to-end Carbon-I design is shown in Figure \ref{fig:pointRFs}. The top two plots correspond to the ARF in green, the middle two plots represent SRF in orange, and the bottom two plots represent the CRF in blue. The right grouping of images are contour plots of the FWHM over the spatial dimension of the FPA in the x-axis and wavelength in the y-axis for each of the three response function requirements. Note that the ARF requirement is uniform over all spatial dimensions and exhibits almost indiscernible increases in FWHM with increasing wavelength due to diffraction in the point spread function (PSF). The SRF FWHM contour plot shows it is uniform over the majority of the wavelengths and field positions, only increasing at the edge field positions. This is a characteristic of the challenging optical design that pushes the image plane size compared to previous class of imaging spectrometer like EMIT. Finally, the CRF FWHM contour plot shows that there is a balance in optical performance between different regions. Note that the edge field positions exhibit high CRF while the center field also exhibits high CRF at the smallest wavelength range. The design optimization balanced the CRF over fields and wavelengths while maintaining excellent SRF and ARF performance. 

\begin{figure*}[!b]
    \begin{subfigure}[b]{0.5\textwidth}
        \centering
        \includegraphics[height=1.9in]{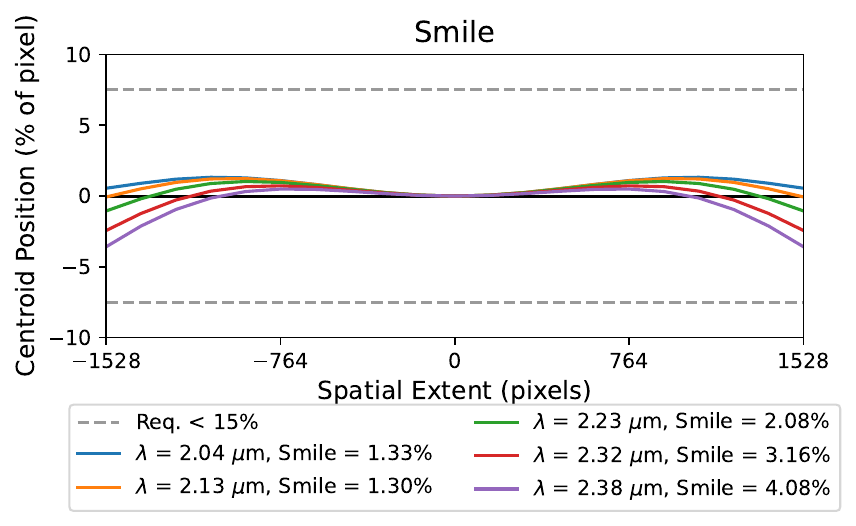}
        \caption{Smile plotted over the spectral range.}
    \end{subfigure}
    \begin{subfigure}[b]{0.5\textwidth}
        \centering
        \includegraphics[height=1.9in]{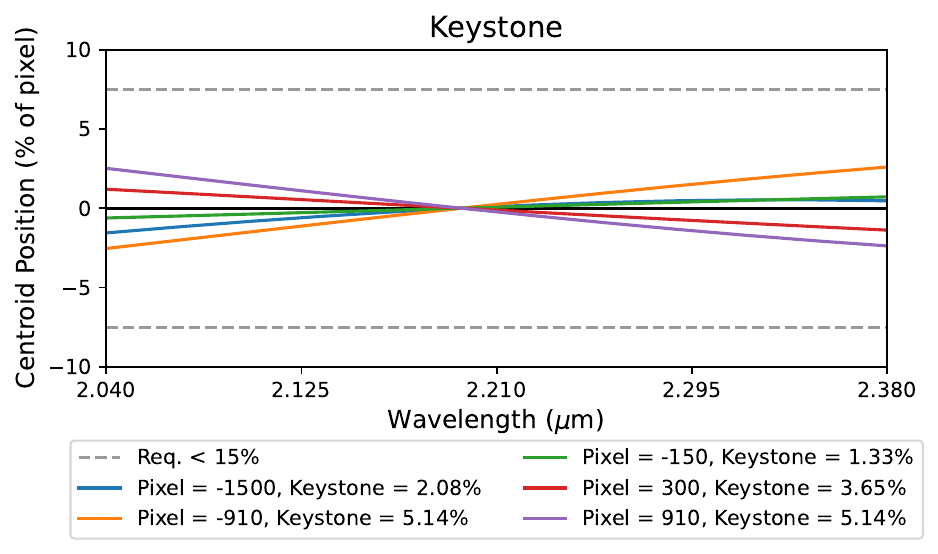}
        \caption{Keystone plotted over the field positions.}
    \end{subfigure}
    \caption{\bf{Centroid Uniformity requirements (a) Smile and (b) Keystone plotted over the wavelengths and fields.}}
    \label{fig:smilekeystone}
\end{figure*}

The plots on the left-hand side show the corresponding shapes and FWHM sizes for the minimum and maximum response functions over the spatial and spectral dimension for ARF, SRF, and CRF. Note the ARF calculation of a convolution of the slit width represented as a rectangular function with motion blur with the same slit width rect function results in a triangular function with rounded edges due to the line spread function (LSF) of the telescope. The slit is twice as large as a pixel so the resulting FWHM is greater than 2 pixels. Also note that there is little difference in the minimum and maximum ARF plots represented in black dotted and black solid lines, respectively, where they are on top of one another. 

\begin{figure*}[!b]
\centering
\includegraphics[width=\textwidth]{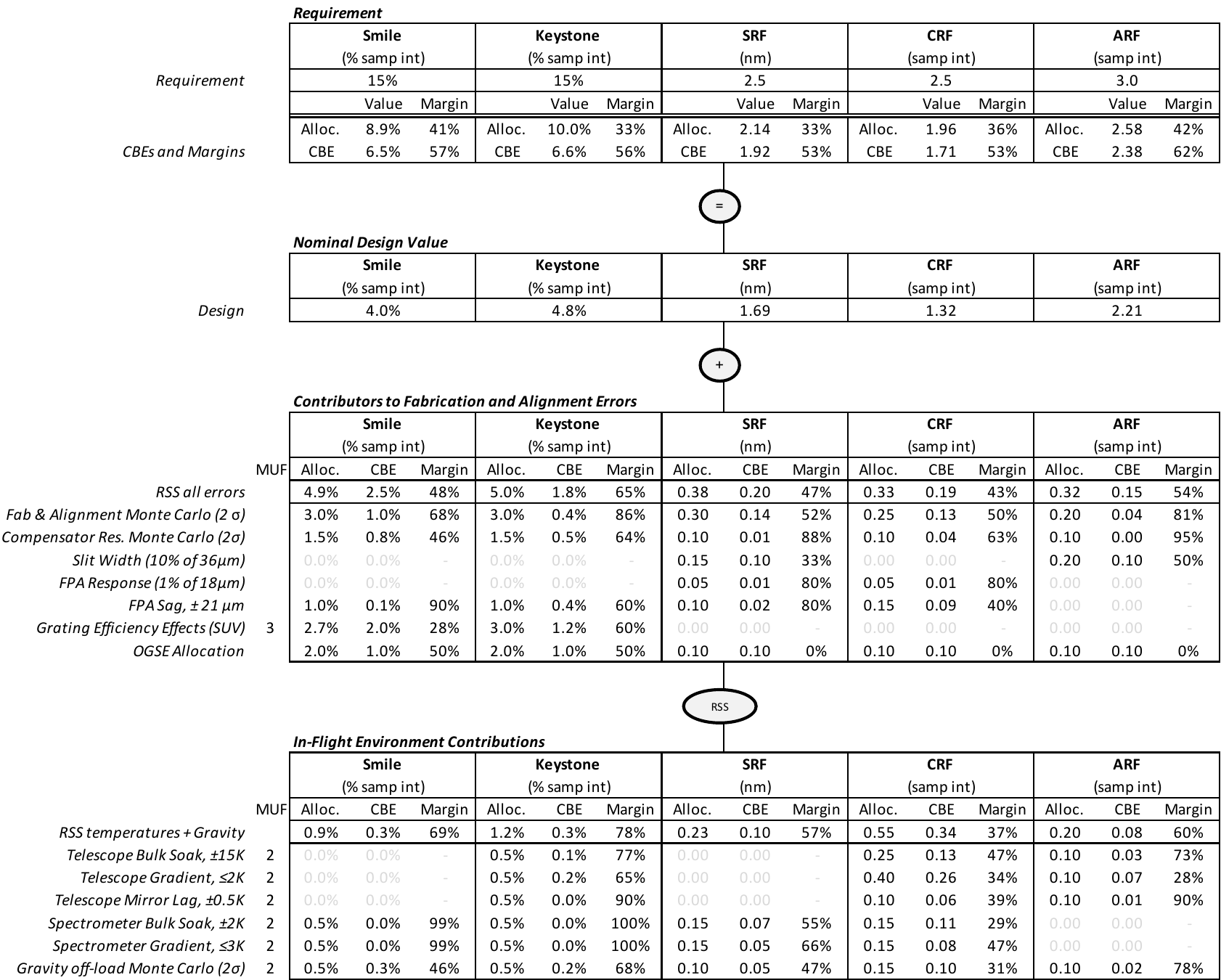}
\caption{\bf{First pass high level error budgets of Smile, Keystone, SRF, CRF, and ARF for the Carbon-I optical design with allocations and current best estimates. Note SRF, CRF, and ARF margins are calculated using equations 1-3.}}
\label{fig:budget}
\end{figure*}

The SRF FWHM cannot be less than twice the slit width converted to spectral units for a spectral sampling of\,0.7\,nm such that SRF $>$\,1.4\,nm. The plot shows that the minimum SRF FWHM is not much larger than the smallest achievable FWHM in the incoherent approximation. As seen in the contour plot, the SRF is close to minimum for the majority of fields and wavelengths. There is an increase in SRF FWHM for edge field positions and represented as the black SRF plot. 

The final plot on the left-hand column shows that the CRF has the smallest starting design value. Based on the incoherent approximation calculation for CRF, it cannot be less than the width of the pixel response. Note that minimum CRF is marginally larger than a pixel width while the maximum CRF FWHM design value is seen at the center field at the lowest wavelength end and the edge field positions. It will be shown in the Error Budget section that the CRF has more contributors to errors so its smallest starting value ensures that all requirements are met simultaneously after errors are considered. The maximum FWHM values are calculated across the field positions and wavelengths and carried for conservative booking in the error budgets. 

The final two optical performance requirements, Smile and Keystone, are plotted for the Carbon-I design in Figure \ref{fig:smilekeystone}. Both requirements are looser than that for SRF and CRF so the design favored the response functions over the centroid uniformity. However, it is shown that the maximum Smile and maximum Keystone are well under the requirement of 15$\%$ to provide healthy margins for errors discussed in the next section.

\section{Error Budgets}
\begin{figure*}[!t]
\centering
\includegraphics[width=0.93\textwidth]{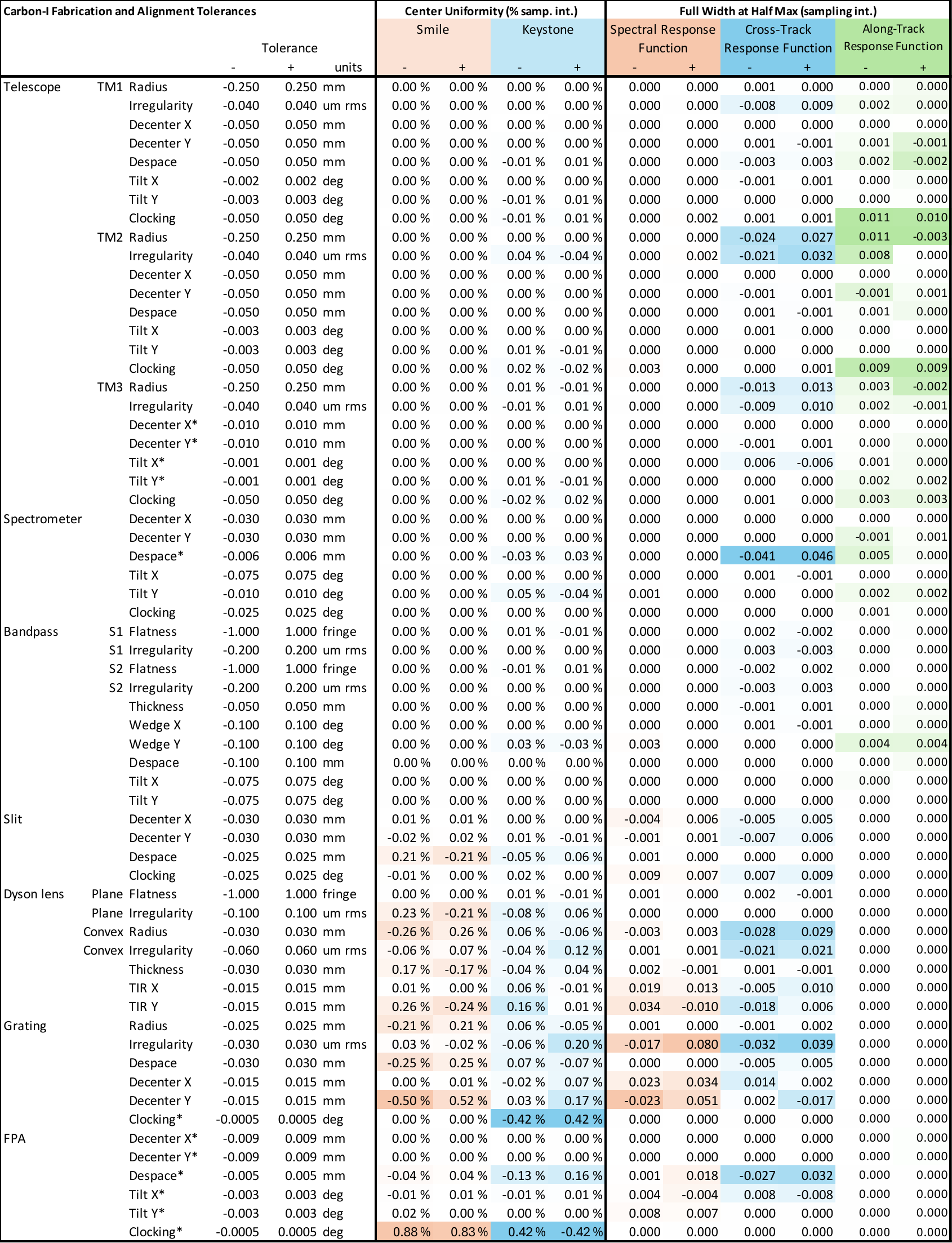}
\caption{\bf{List of fabrication and alignment tolerances and compensators (indicated with $\ast$) used to assess the as-built term in the high level error budgets. The leftmost grouping of text lists the tolerance parameter and values. The right side of the table lists the 5 optical performance requirements with their corresponding impact due to the tolerance value. A cell highlighted in color indicates that the tolerance has an impact on the requirement. The worst offending tolerances are represented by a higher saturation in color.}}
\label{fig:tolerances}
\end{figure*}

To assess the feasibility of the Carbon-I optical design implementation, first pass error budgets shown in Figure \ref{fig:budget} are developed to include errors such as fabrication, alignment, test equipment called optical ground support equipment (OGSE), FPA flatness, and in-flight environments due to thermal, and gravity off-load. Carbon-I utilizes tolerancing methods and tools developed for previous instruments that assess all 5 optical performance requirements simultaneously \cite{Moore10.1117/12.2568857}. Errors for fabrication and alignment define tolerances for procurement of optical elements, optomechanical mounting stresses, and placement tolerances and compensators for alignment activities. 

Preliminary Carbon-I tolerances are in line with typical requirements for fabrication, mounting, and alignment of previous imaging spectrometers. The tolerances and their impacts on optical requirements are listed in Figure \ref{fig:tolerances}. The left-most group of columns lists the tolerance parameters for each component in the optical system. The five right-most columns correspond to the five imaging spectrometer optical performance requirements. These columns carry the change in the requirement value or the requirement sensitivity to tolerances through the use of color saturation. A cell with color indicates that the tolerance is sensitive and a high saturation in color indicates the tolerance has a large impact on the optical performance requirement. As an example, the TM2 Radius tolerance of $\pm$ 0.25\,mm shows it will increase the CRF FWHM by 0.027 pixels and the ARF FWHM by 0.011 pixels. Note that it has no impact on smile and SRF FWHM as these are spectrometer specific requirements. With this representation it is shown that the CRF FWHM has more contributors to errors than other requirements, which requires a smaller CRF starting value. The number of error contributors for SRF is smaller and allows for the key requirement for GHG characterization to be comfortably achieved. Highlighted in this table are also the compensator resolutions with lockdown errors that are indicated with the use of an asterisk. Any parameter that has an asterisk behind the name is used during alignment activities to compensate for placement or fabrication errors. 

\begin{figure*}[!t]
\centering
\includegraphics[width=\textwidth]{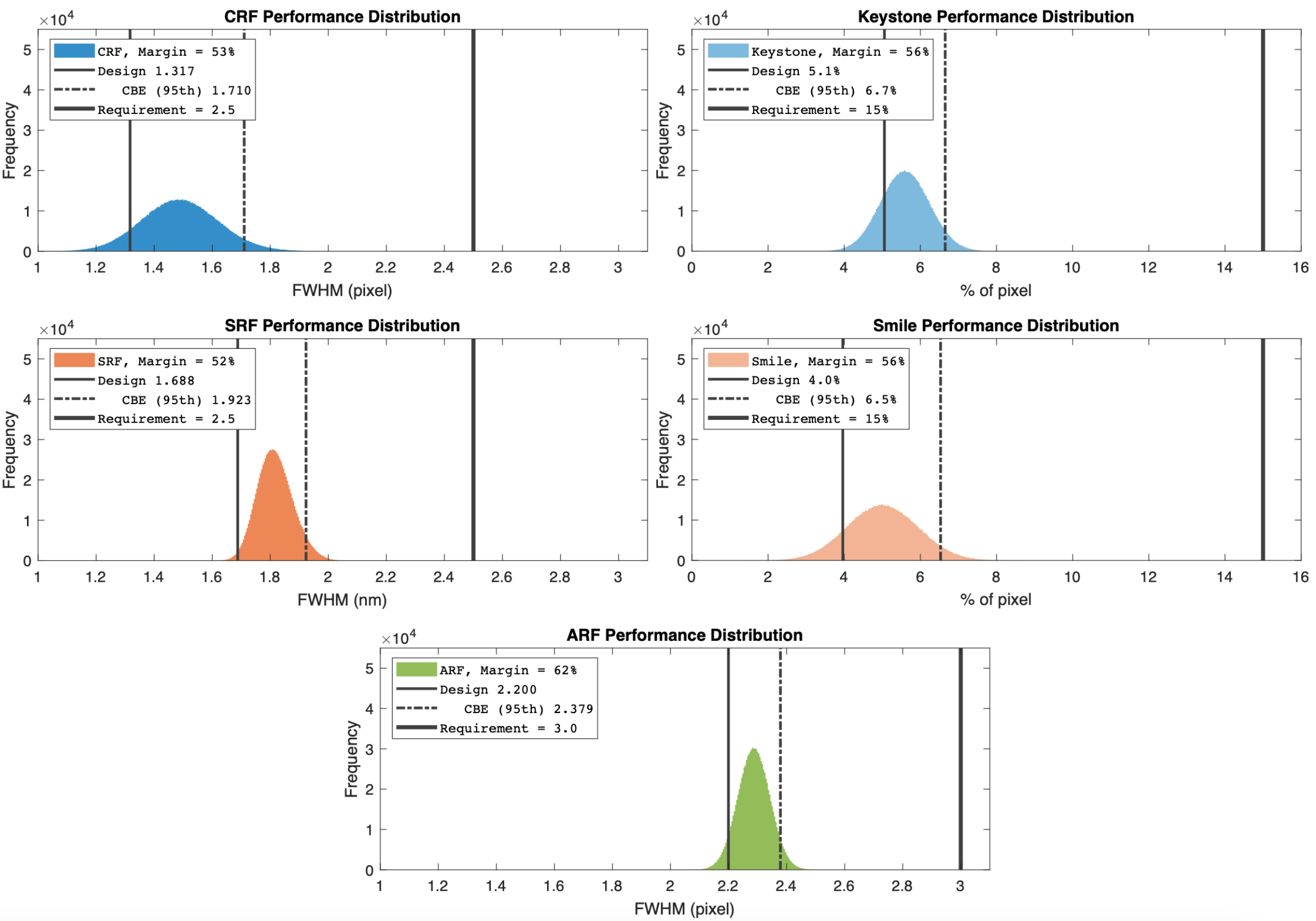}
\caption{\bf{Histograms of all Monte-Carlo analyses along with a roll-up of all errors due to in-flight environments are presented for the 5 optical performance metrics.}}
\label{fig:histograms}
\end{figure*}

The fabrication and alignment tolerances in Figure \ref{fig:tolerances} were varied through a Monte Carlo analysis to understand the 2-$\sigma$ deviation of errors as the CBE. Other fabrication and alignment errors include the effect of slit width variation on ARF and SRF and the detector response function uncertainty for SRF and CRF requirements. A large contributor to the fabrication and alignment errors is the FPA surface sag. With a large FPA format, it is expected that the surface sag can deviate by approximately 20\,$\mathrm{\mu}$m, thereby causing defocus across the image plane of the spectrometer. We also reserve some of the margin allocation for the test equipment called Optical Ground Support Equipment (OGSE) that will probe the instrument to diagnose misalignments and calibrate the instrument. These errors are presented in the high-level error budgets presented in Figure \ref{fig:budget} in the “Contributors to Fabrication and Alignment Errors” box. The box at the bottom shows allocations to the In-flight environments that the instrument can see while operating in space.

Zemax thermal modeling was used along with baseline material choices to vary optical prescription parameters per material’s coefficient of thermal expansion (CTE) to understand the impacts of in-flight thermal environments. The environments calculated for the error budgets include bulk soak variation of the telescope bench, gradients along the optical axis of the telescope bench, temperature lag of mirrors relative to the structure due to bipod mount design, the spectrometer bulk soak variation, and spectrometer gradient across the optical axis. Most of the structure is assumed to be aluminum at this formulation stage of the project. Telescope mirrors are aluminum, the Dyson lens is fused silica, and the baseline material choice for the grating substrate is N-BK7. 

\subsection{Telescope Bulk Soak Range}

The telescope operates at room temperature and is baselined to be all aluminum. In the case where all telescope elements such as the structure and mirrors share the same CTE, an isotropic thermal soak will result in no change in optical performance. However, there is an uncertainty in the CTE of aluminum if all parts come from different lots of material. We carry an aluminum uncertainty of $\pm$1$\%$ to find the maximum variation in telescope bulk temperatures. The procurement plan for the telescope structure involves fabricating all bench parts out of the same lot of aluminum such that it is assumed all structure parts will have the same CTE. We evaluated 14 combinations of maximum CTE variations across between four components to include the structure, telescope mirror 1 (TM1), telescope mirror 2 (TM2), and telescope mirror 3 (TM3) in the simple thermal model. As an example, one configuration assigned CTE variation of TM1, TM2, and TM3 to be +1$\%$ while the bench was held to a CTE of -1$\%$. The mirror radius of curvatures, the freeform Zernike coefficients, and the Zernike normalization radius varied with the mirror CTE while the bench structure CTE varied the TM1-to-TM2, TM2-to-TM3, and TM3-to-slit distances. Each of the 14 configurations were evaluated through $\pm$20 K to understand to bound the worst-case configuration to carry in the error budgets. Through this exercise, it was found the maximum temperature variation in telescope bulk soak due to this aluminum CTE uncertainty is $\pm$15\,K. The change in performance due to $\pm$15\,K bulk soak change is carried in the error budgets with a model uncertainty factor (MUF) of 2. The MUF is conservative as these are simplified thermal models that will later be replaced by proper Structural (FEM) Element Model analysis.

\subsection{Telescope Gradient}

A similar Zemax thermal model was developed to assess the error due to a gradient across the telescope bench. The worst-case orientation of the gradient is along the optical axis of the telescope for the optical performance metrics. For this environment, the CTE is assumed to be the same for all parameters in the telescope optical prescription. The Zemax thermal model applies a gradient by maintaining the bench at nominal temperature as the TM2 mirror is cooled and the TM1 and TM3 mirrors are heated or vice versa. Gradients of $\pm$5 K were evaluated to show that the error budgets could accommodate a maximum gradient of 2K. 

\subsection{Telescope Mirror Temperature Lag}

Due to the bipod mirror mount design, there is limited conductance from the telescope bench to the mirrors that results in the mirrors lagging in temperature with respect to the structure. This effect was seen through Structural Thermal Optical Performance (STOP) modeling of the EMIT instrument and is expected to exist for similarly mounted telescope mirrors like Carbon-I. To model this effect, the nominal CTE of aluminum is applied to all telescope prescription parameters. The bench was held at nominal operating temperature while the TM1, TM2, and TM3 were assumed to vary by the same temperature. The preliminary temperature variation that is carried in the error budgets is $\pm$0.5\,K. 

\subsection{Spectrometer Bulk Soak Range }

The operating temperature of the spectrometer is 240K to reduce the thermal emission. The Carbon-I spectrometer is not an athermalized design. With the baseline material choices where the Dyson lens is fused silica, the structure parts are aluminum, and the grating is N-BK7, there will be large changes in optical performance due to the CTE mismatches. While an aluminum grating substrate would match the structure CTE to produce a less sensitive spectrometer, the grating fabrication is better suited to glass substrates and thus chosen as the baseline for the Carbon-I spectrometer. The Zemax thermal model was set to vary the Dyson lens radius of curvature, thickness, and even asphere terms with the CTE of fused silica. The CTE of aluminum was applied to the distances between the Dyson lens and grating, slit-to-Dyson lens, and FPA-to-Dyson lens. The N-BK7 CTE was assigned to the grating radius of curvature. Temperature variations of $\pm$10\,K were evaluated and shown that the error budgets can tolerate $\pm$2\,K bulk soak change. 

\begin{figure*}[!t]
\centering
\includegraphics[width=\textwidth]{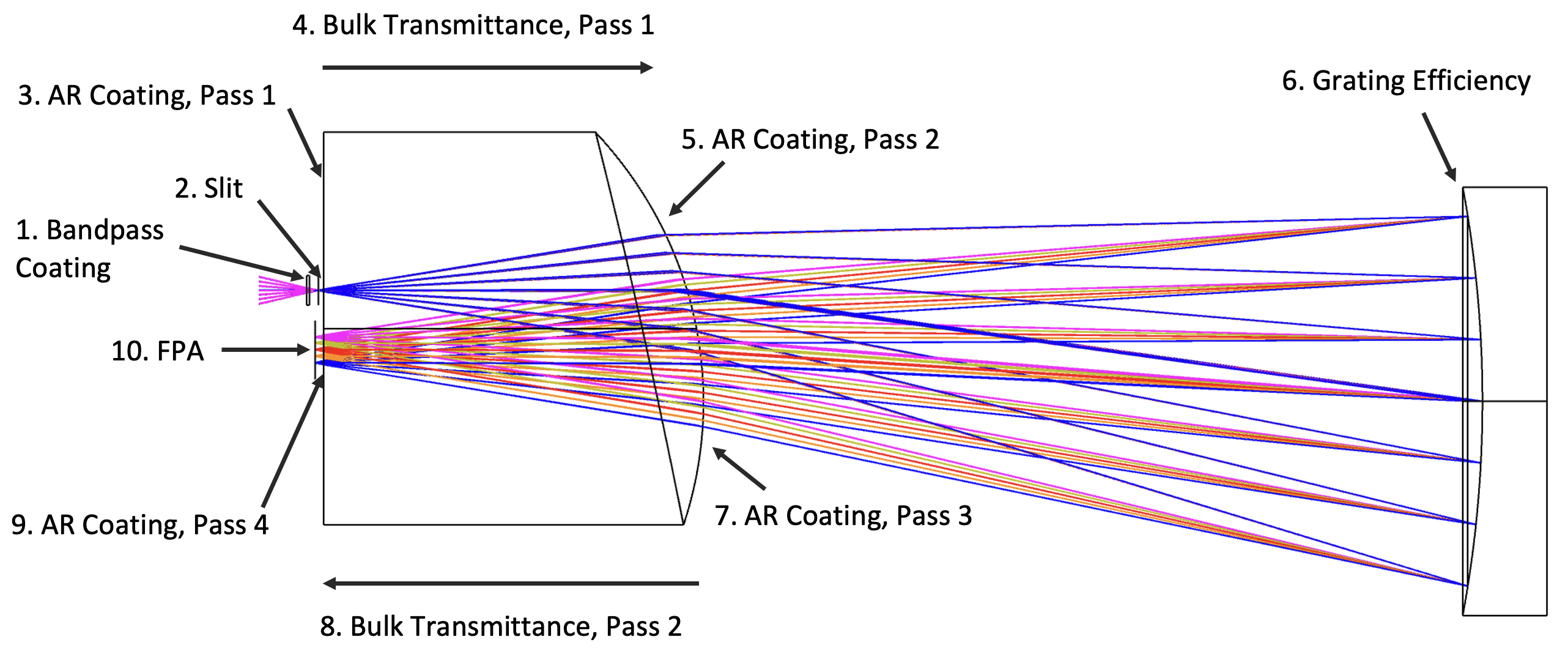}
\caption{\bf{Carbon-I Spectrometer Ray Path. The photon ray path is as follows: Photons from the telescope travel left to right to first (1) pass through a bandpass filter for stray light mitigation before entering the (2) spectrometer through the slit. After the slit, (3) the light passes through the plano-convex surface coated with an antireflection (AR coating). (4) Photons now make their first pass through the Dyson lens traveling left to right where they can experience bulk losses from material absorption and scatter from bubbles and inclusions. (5) The photons then encounter the AR coated convex surface of the Dyson lens to refract toward the grating. (6) The photons travel in free-space to encounter the grating with a gold coated, blazed grating to reflect (photons now travel right to left in the figure) and disperse light back toward the Dyson lens. (7) Photons encounter the AR coated convex surface of the Dyson lens a second time. (8) The phones travel right to left now make their second pass through the lens to experience bulk losses to finally reach the AR coated plano-surface. (9) The light passes through the AR-coated plano-surface to finally reach the (10) focal plane where the FPA QE is applied.}}
\label{fig:raypath}
\end{figure*}

\subsection{Spectrometer Gradient}

Using the same baseline material choices and CTEs defined in the spectrometer bulk soak range calculation, the gradients are applied along the optical axis of the spectrometer. Previous imaging spectrometer designs such as EMIT use a thermal strap on the spectrometer body located either near the FPA or the grating of the spectrometer. To evaluate the impact of the gradients, the slit and FPA end of the spectrometer was held nominal temperatures while the Dyson lens would change by 1/4 of the gradient, the bench would change by 1/2 of the gradient, and the grating would change by the full gradient. This analysis showed that the gradient that can be accommodated in the error budgets is $<$3K across the spectrometer and a similar MUF of 2 is applied due to the simplicity of the modeling.

\subsection{Gravity Release}

The final environment to book the error budgets is due to gravity off-load as the instrument is aligned with gravity and launched to space. Optical elements and structure will shift from their aligned configuration. To provide an allocation and guidance to the optomechanical design, small uncompensated errors were chosen as tolerances for the optical prescription. They were varied through a Monte Carlo analysis and the 2-sigma is carried in the error budgets. The individual tolerances can provide component level guidance during the first pass mounting designs with the use of breakout FEMs. However, once the system level FEM is developed, the load specified by the thermal requirements and 0-g will be applied and mapped to the Zemax optical prescription to understand errors in the optical performance. When FEMs are used to evaluate performance, the MUF will be reduced. 

The error budgets contain allocations for all known errors from previous programs where there is a minimum of 33$\%$ of unallocated margin for all requirements for other errors that may arise through the detailed design phase of the instrument. The roll-up of the current best estimates for each error shows a minimum of 53$\%$ margin across all optical performance requirements. The definition of the fabrication and alignment tolerances give initial requirements on the optomechanical structure and the in-flight environment calculations provide initial thermal requirements to maintain optical performance while operating. This roll-up of errors and statistical analysis of the point design shows that the design meets the instrument requirements as indicated in the histograms in Figure \ref{fig:histograms}. The statistical analysis shows that all cases evaluated meet the optical performance requirements simultaneously and thus the program is carrying sufficient margins to move into the next design phase.

\section{Optical Throughput}

Dyson imaging spectrometers experiences throughput losses due to the double-pass operation of the spectrometer as illustrated in Figure \ref{fig:raypath}. However, the optical efficiency of Carbon-I is higher than that of previous imaging spectrometers that operate over the 380-2500\,nm wavelength range. As previously mentioned, instruments such as EMIT and SBG VSWIR require shaped groove gratings to distribute the grating efficiency at the short-wave infrared to the visible wavelengths resulting in reduced efficiency per wavelength. Carbon-I samples a much smaller subset of the spectral range that can utilize a simpler blazed grating design optimized for the Carbon-I center wavelength. This smaller spectral range also enables the use of simple but highly effective anti-reflection (AR) coatings on the transmissive optics. The high dispersion of this spectrometer to achieve the\,0.7\,nm spectral sampling also pushes all other diffraction orders away from the FPA. An order sorting filter is not required for the Carbon-I design. Instead, we include a bandpass filter in front of the spectrometer slit for stray light control, which allows better straylight control compared to EMIT. 

\begin{figure*}[!t]
\centering
\includegraphics[width=\textwidth]{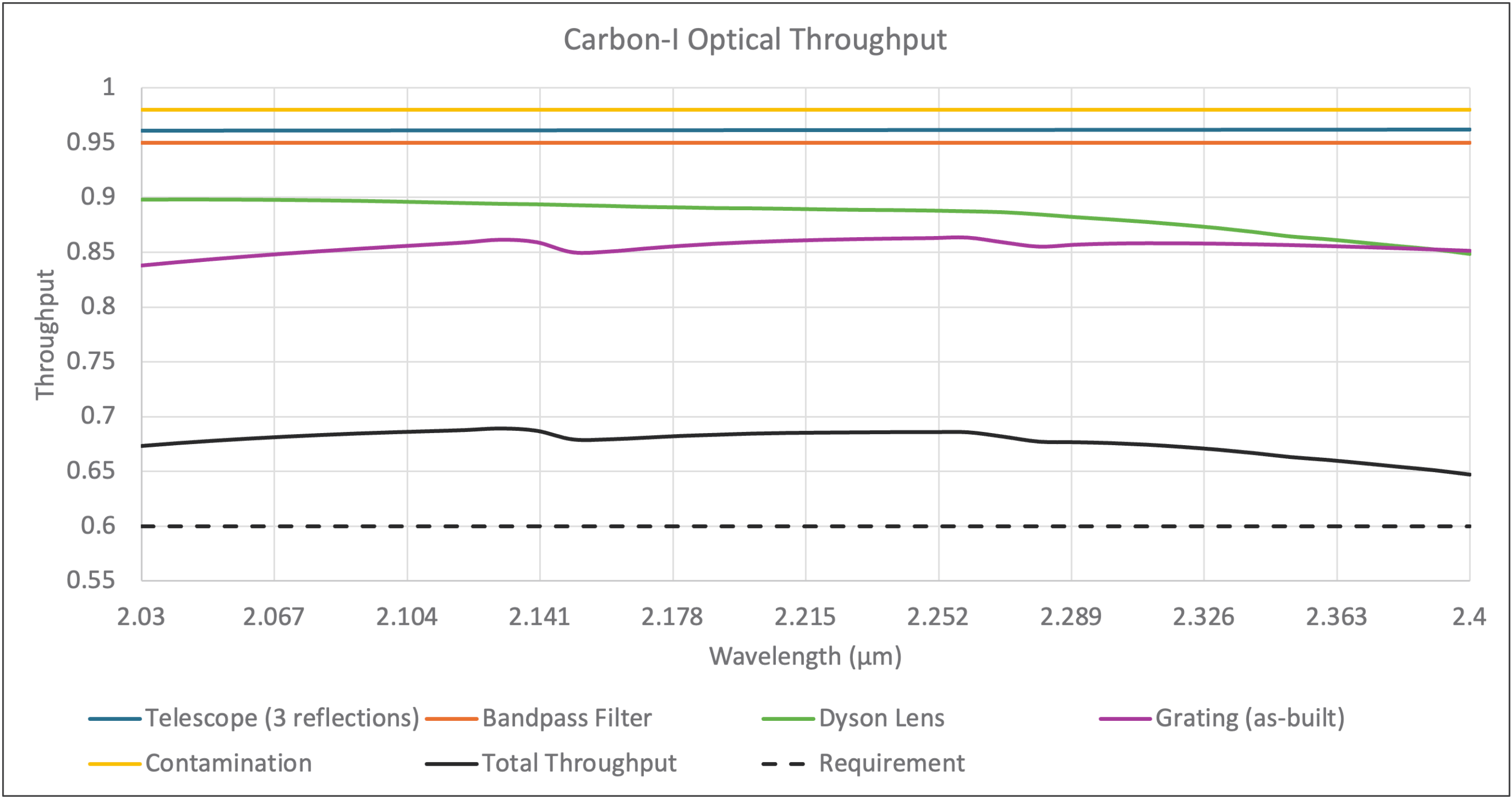}
\caption{\bf{Carbon-I Optical Throughput. The Carbon-I optical throughput is the product of the efficiencies of the telescope, bandpass filter, Dyson lens, grating, and 2$\%$ contamination allowance.}}
\label{fig:throughput}
\end{figure*}

The Carbon-I optical throughput is affected by four main optical components: the telescope, the bandpass filter, the Dyson lens, and the grating. We also add a fifth factor for contamination accumulated by the instrument end of life. The optical system throughput is calculated by multiplying the transmittance of each element and shown in Figure \ref{fig:throughput}.

\subsection{Telescope}

The telescope transmittance term accounts for three reflections for protected gold mirror surfaces. These surface reflectances are calculated for optical prescription angle of incidences for gold with a thin Silicon Oxide layer as the protective topcoat. Each reflection has an efficiency of approximately 0.982 at our lowest wavelength. We account for a 0.5$\%$ uncertainty in the mirror reflectance, verified using multiple measurements and models of protected gold coatings at different angles of incidence. Accounting for three reflections, the transmittance of the telescope is 0.961 at the low wavelength end. The slit diffraction loss at the interface between the telescope and spectrometer is negligible because our Airy disk diameter of 13 $\mathrm{\mu}$m is much smaller than the 36 $\mathrm{\mu}$m slit width. 

\subsection{Bandpass Filter}

The bandpass filter placed in front of the spectrometer slit at a tilted angle of 1 degree that will block all wavelengths outside of the Carbon-I spectral range. The bandpass filter is coated sapphire with similar specifications as the order sorting filter zones required for VSWIR imaging spectrometers. We carry a 0.95 transmittance requirement for the bandpass filter along with OD 4 blocking for the wavelengths outside of the Carbon-I spectral range. 

\subsection{Dyson Lens}

The Dyson spectrometer works in double pass such that the optical path interacts with the Dyson lens twice. The Dyson lens efficiency term includes four surface interfaces and two passes through the bulk transmittance of the fused silica lens with scatter due to bubbles and inclusions of 0.25$\%$ as shown in Figure \ref{fig:raypath}. All four surfaces are AR-coated with a reflectance calculated to be approximately 0.3$\%$ each using Angus Macleod’s coating software. This AR-coating design features four alternating layers of common coating materials, Titanium Oxide and Silicon Dioxide. The layer thicknesses are optimized through the coating software tool over the Carbon-I wavelength range and F/$\#$. We derate these predictions by 0.3$\%$ which corresponds to the 95th percentile reported at the long wavelength for fabrication uncertainties in layer thicknesses. Finally, we account for bulk transmittance of the optical path along each direction through the lens. We take bulk transmittance values from vendor data on the Heraeus Suprasil 3001 fused silica glass product for our 200 mm path length \cite{HerausConamic2023,Covantics2024}. The fused silica bulk transmittance degrades at the longer wavelengths resulting in Dyson lens minimum transmittance of 0.848. 

\subsection{Grating}

The Carbon-I grating design builds upon the Jet Propulsion Laboratory Microdevices Laboratory experience fabricating concave gratings \cite{Wilson10.1117/12.510204}. The Carbon-I grating efficiency is simulated using PCGrate-SX software that performs a full-wave polarized electromagnetic solution of Maxwell’s equations using an integral equation method. For the grating design, we specify the grating period, groove shape (blaze and step angles), angle of incidence, and gold metallic coating. The software then calculates the absolute efficiencies of all diffraction orders at all specified wavelengths. The complex refractive index of the gold coating is included where the software accurately predicts absorption losses and grating resonances called Wood’s anomalies. This software has been used for two decades and its predictions match optically measured grating efficiencies. Figure \ref{fig:grating} shows atomic force microscopy measurements of a test grating’s blaze. The AFM profiles were imported to PCGrate to simulate the efficiency to inform better estimates of the as-fabricated grating efficiency. The plot shown in Figure \ref{fig:throughput} incorporates the as-built grating efficiency from the first test grating fabrication run. The as-built grating efficiency includes fabrication errors due to grating depth, grating roughness, and groove corner rounding. Improvements are expected for future test grating fabrication runs as the team tunes the writing parameters. Nevertheless, the initial grating fabrication errors generate a conservative optical throughput that still meets the requirement of $\ge$0.6 across the wavelength range. 

\subsection{Contamination}

The end-of-life contamination value was taken from the EMIT mission simulations. This instrument has similar exposed surface areas and will be constructed with similar contamination control protocols. Note that EMIT is considered a worst case condition as it operates on the International Space Station. It is expected that Carbon-I will have lower background contamination environment since the spacecraft components see bake-out prior to instrument integration. Molecular contamination was derived from the absorptivity of contamination outgassed during a previous thermal vacuum test, and then scaled with wavelength through the Beer-Lambert thin slab model. 

All contributors and final optical system throughput are presented in Figure \ref{fig:throughput}. These calculations do not include the quantum efficiency (QE) of the detector or optical reflection at the detector surface, which is allocated under a separate QE allocation outside of the optical throughput allocation. Recent measurements of modern CHROMA detectors for instruments like CPM indicate a QE across the entire detector around 0.85. The Carbon-I Optical throughput is calculated to be within the 0.65 – 0.69 including first test grating efficiency data, which is better than the requirement of 0.60. 

\section{Stray Light Assumptions}
\begin{figure*}[!t]
    \begin{subfigure}[b]{0.5\textwidth}
        \centering
        \includegraphics[height=2.7in]{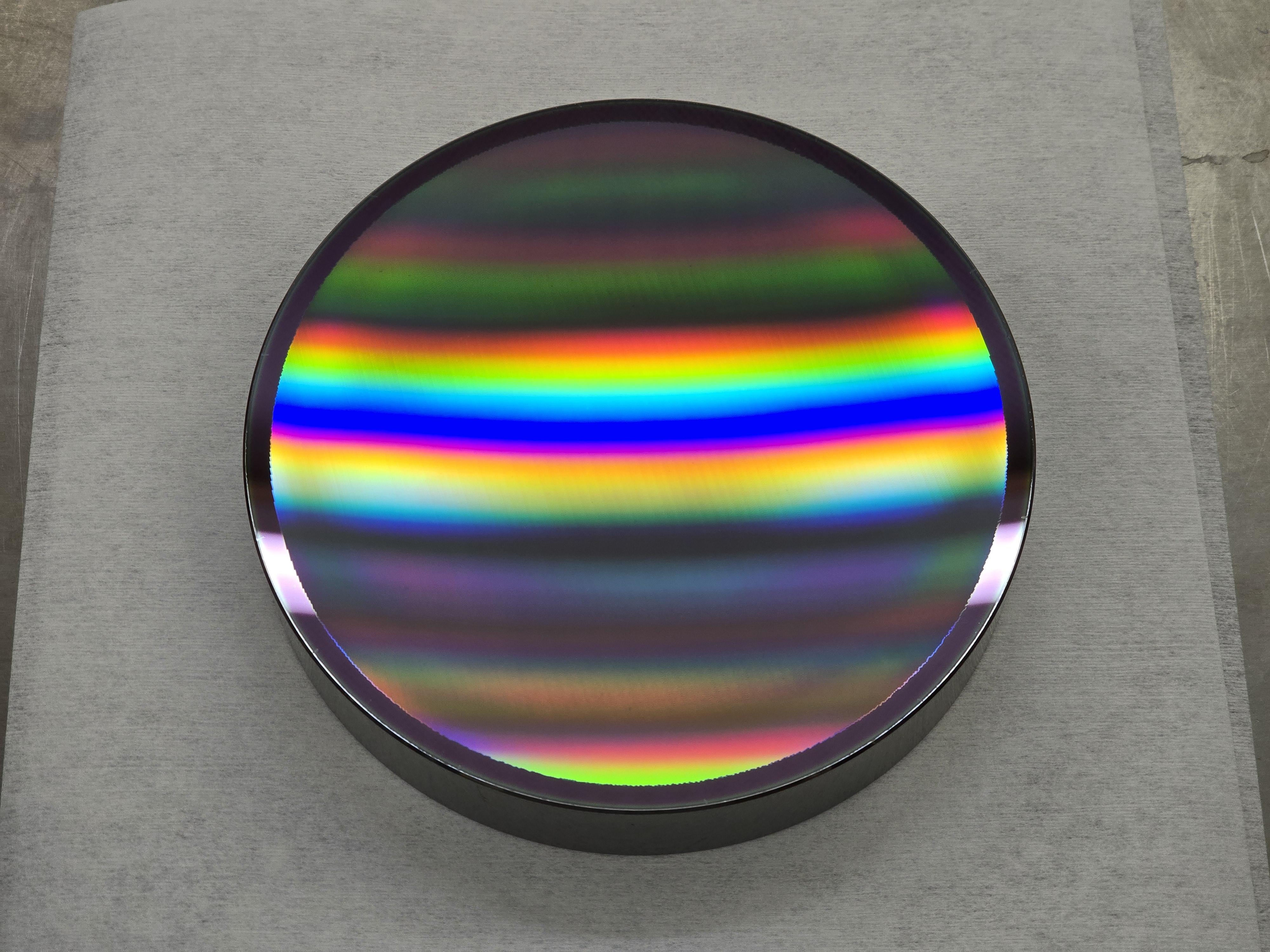}
        \caption{Carbon-I test grating  on a concave spherical glass substrate.}
    \end{subfigure}
    \begin{subfigure}[b]{0.5\textwidth}
        \centering
        \includegraphics[height=2.7in]{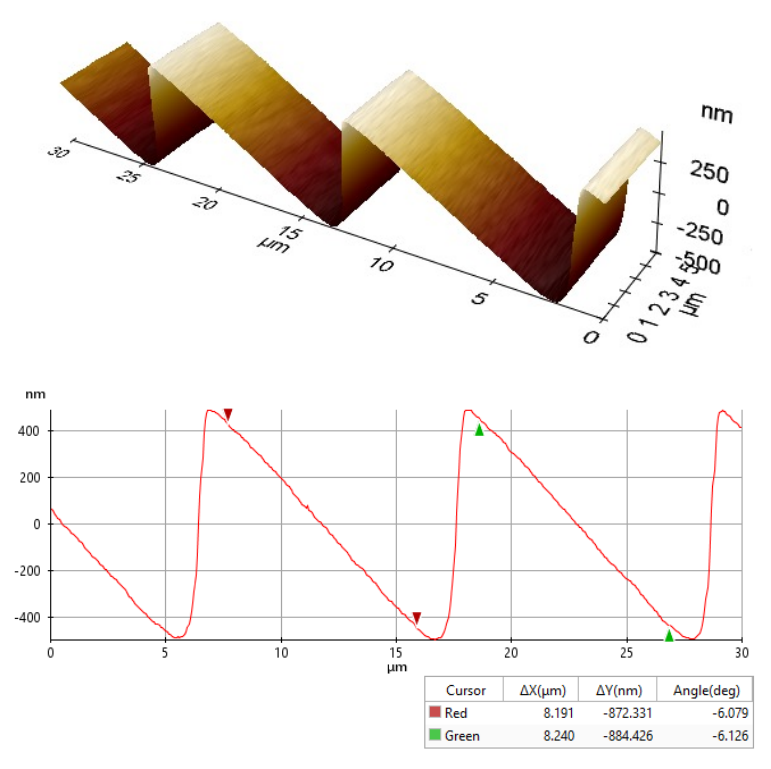}
        \caption{Atomic Force Microscope (AFM) scan of the Carbon-I test grating.}
    \end{subfigure}
    \caption{Carbon-I Grating development in (a) test grating on a concave spherical glass substrates with (b) Atomic Force Microscope scan of the blazed grooves.}
    \label{fig:grating}
\end{figure*}

As mentioned in the Telescope design discussion, the SBG VSWIR F/1.8 telescope is stopped down to F/2.2 to match the Carbon-I spectrometer design. This allows the system to make use of efficient vanes for extended field stray light mitigation. The telescope vanes eliminate the direct illumination of the M3 telescope mirror and effectively shadow the slit of the spectrometer. The baseline design also incorporates black Z306 paint for all structure parts of the telescope and spectrometer. 

\begin{figure}[!b]
\centering
\includegraphics[width=0.47\textwidth]{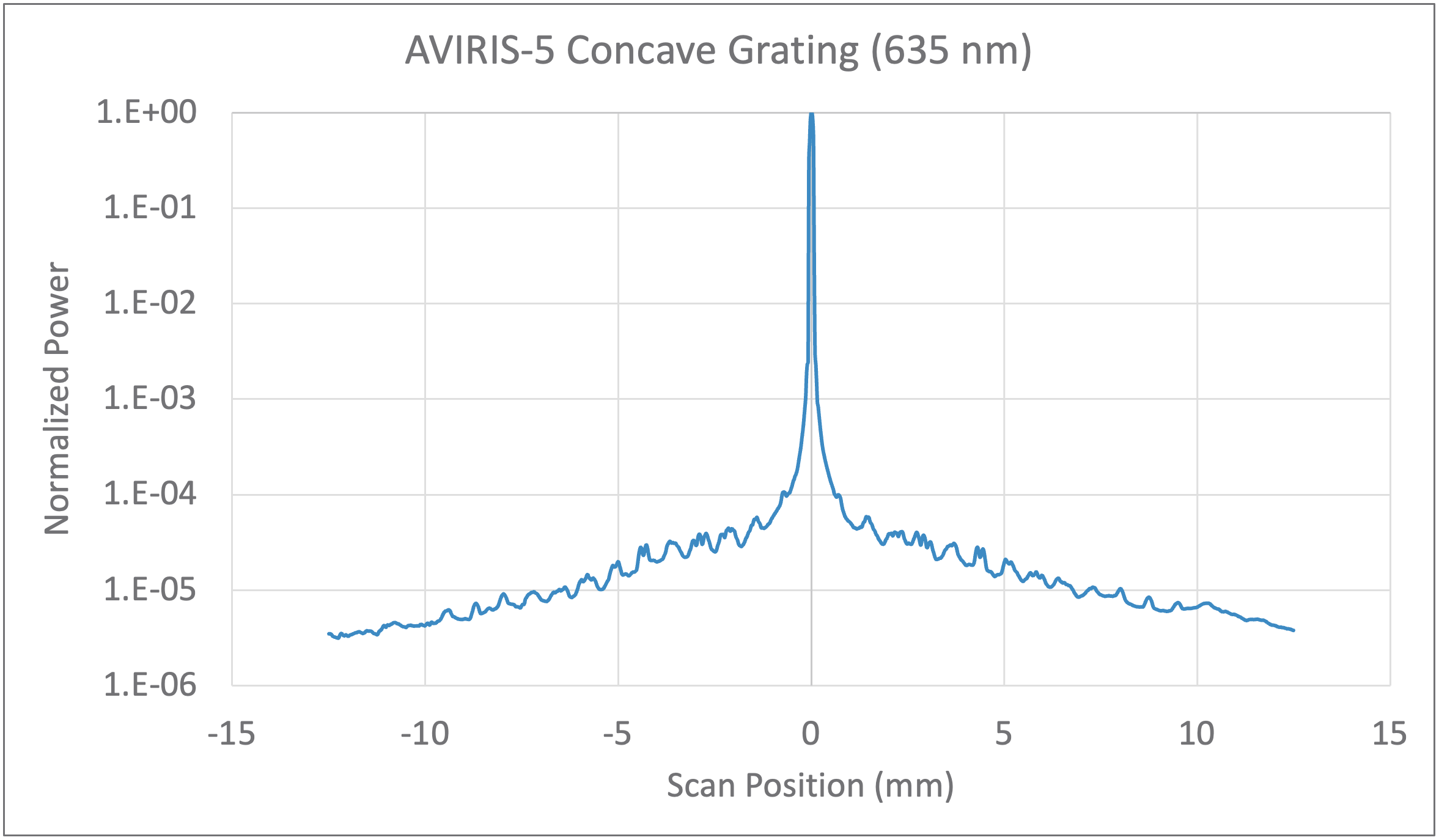}
\caption{\bf{Grating scatter measurement of the AVIRIS-5 grating written with same process as Carbon-I grating on concave substrate.}}
\label{fig:gratingscatter}
\end{figure}

Stray light analyses performed by Photon Engineering on the EMIT program showed that the primary contribution to the extended field stray light is scatter of the telescope mirror surface. With the assumption that the telescope is well baffled, the mirrors are the only surfaces that are both illuminated and critical, i.e. surfaces viewed directly by the FPA, in the system where they will be the main contributors to stray light. As the Carbon-I wavelength is much longer than the shortest EMIT wavelengths, the surface roughness of the telescope mirrors can be loosened. The current baseline carries a surface roughness requirement of 40 Angstroms rms compared to 20 Angstroms rms for EMIT. The total integrated scatter (TIS) is represented by 

\begin{equation}
    \textnormal{TIS}(\lambda) = \left(\frac{2\pi\Delta n \sigma_{rms}}{\lambda}\right)^2,
\end{equation}

where $\Delta$n is 2 in the case of a mirror, $\sigma_{rms}$ is the surface roughness, and $\lambda$ is the wavelength \cite{Fest2013}. At 2000 nm and 40 Angstroms rms, the TIS is $\sim$0.063$\%$ per mirror in the telescope. In a conservative estimate, let’s assume that the TIS from each mirror contributes to the overall scatter such that the maximum extended field stray light is 0.19$\%$, which is much lower than the 10$\%$ extended field stray light requirement.

The in-field scatter for grating-based imaging spectrometers is dominated by the grating scatter. Other contributors include scatter from roughness of optical elements, bulk scatter of transmissive optics, cleanliness, and ghost reflections. Previous analysis done for MISE, EMIT, and CPM show that the grating will overwhelm the scatter contributions from all other sources. Due to the high dispersion of the Carbon-I grating, there are no overlapping diffraction orders on the detector. This results in a single peak on the detector. Figure \ref{fig:gratingscatter} shows the measured grating scatter for a grating fabricated by the same process needed for the Carbon-I grating on a spherical substrate and results in excellent stray light performance resulting in tails lower than $10^{-5}$. The Carbon-I grating scatter performance is expected to be similar and has the advantage of operating at longer wavelengths that will exhibit less scatter.

The team continues to develop test gratings for the proposal. The first grating fabrication test is shown in Figure \ref{fig:grating}. The Carbon-I dispersion is large as can be seen with the diffracted colors on the grating. Atomic Force Microscope (AFM) scans were taken and used to inform the fabrication process and estimate grating efficiency.

\section{Conclusions}

The Carbon-I imaging spectrometer for the identification of greenhouse gases CH$_4$ and CO$_2$ emission hotspots is presented. Carbon-I will differentiate the sources and composition of emissions at the city block level.  For the detection of GHG, the instrument is required to image at high spatial resolutions and sample the spectra finely over short-wave infrared wavelengths. Straylight needs to be minimized and the spectral response function should be stable and well characterized. The Carbon-I design achieves 30\,m ground sampling distance with 0.7\,nm spectral sampling. 

The proposed Carbon-I optical design was evaluated across all first-order design requirements and the five key and driving optical performance requirements: Along-track response function (ARF), Spectral Response Function (SRF), Cross-track Response Function (CRF), Smile, and Keystone. With JPL tools for evaluating imaging spectrometer performance, preliminary tolerancing and error budgets show the proposed design meets the optical performance requirements with ample margin. The errors accounted for in the error budgets are consistent with PDR level budgeting for JPL imaging spectrometers. This exercise also defined initial requirements for the optomechanical structure mounts, the alignment, and the thermal requirements across the optical bench. The thermal requirements were assessed through simple CTE calculations using baseline material choices and varying optical prescription parameters with temperature.

The Optical Throughput is also presented based on coating and material choices of the optical elements. The Optical Throughput relies on five factors: the telescope mirror reflectance, the bandpass filter, Dyson lens transmission, grating efficiency, and end-of-life contamination. Telescope mirrors are baselined to be protected aluminum. The bandpass filter is booked as a transmittance requirement for throughput calculations. The Dyson lens is AR-coated and the bulk transmittance follows Heraeus Suprasil 3001 material data. The grating is simplified from VSWIR imaging spectrometers since it operates over a much narrower spectral range. The grating is a sawtooth design with grating efficiency maximized at the center wavelength of the spectral range to provide high grating throughput across all wavelengths. Lastly, conservative contamination contribution is accounted for based on EMIT practices. The optical throughput of the Carbon-I design presented in this paper includes deratings to account for fabrication of coatings for mirrors, bandpass filter, AR coating for the Dyson lens and includes the grating efficiency from Carbon-I's first test grating to show it meets the 0.6 optical throughput requirement across the entire wavelength range. 

The Carbon-I design lends itself to stray light mitigation from extended source. The telescope allows for effective baffles and vanes to shadow the TM3 mirror within the telescope and shadow the slit of the spectrometer. There are no direct stray paths from the external world to the slit of the instrument. Previous extended field stray light analysis done for EMIT and CPM show that the mirror scatter is the largest contributor. Using this assumption with a conservative calculation of the total integrated scatter for each mirror at the Carbon-I wavelength for the baseline mirror roughness, we show that we can expect the scatter contribution to be $<$0.5$\%$. The in-field stray light for previous instruments has shown to be dominated by the grating scatter. It is expected that the Carbon-I grating will have similar scatter performance as previous imaging spectrometers such as AVIRIS-5 that displays tails below $10^{-5}$.  

The work presented in this paper offers engineering solutions of the proposed design for Carbon-I to meet the optical performance requirements necessary for GHG identification and characterization. The team is progressing toward submitting the second phase of the proposal and developing a testbed imaging spectrometer that will use a testbed grating to measure spectra over our wavelength range and further improve the grating efficiency and scatter. 

\acknowledgments
The work detailed here was carried out at the Jet Propulsion Laboratory, California Institute of Technology, under contract with the National Aeronautics and Space Administration (80NM0018D0004). 

\bibliography{carbon-i}   
\bibliographystyle{IEEEtran}

\thebiography
\begin{biographywithpic}
{Christine L. Bradley}{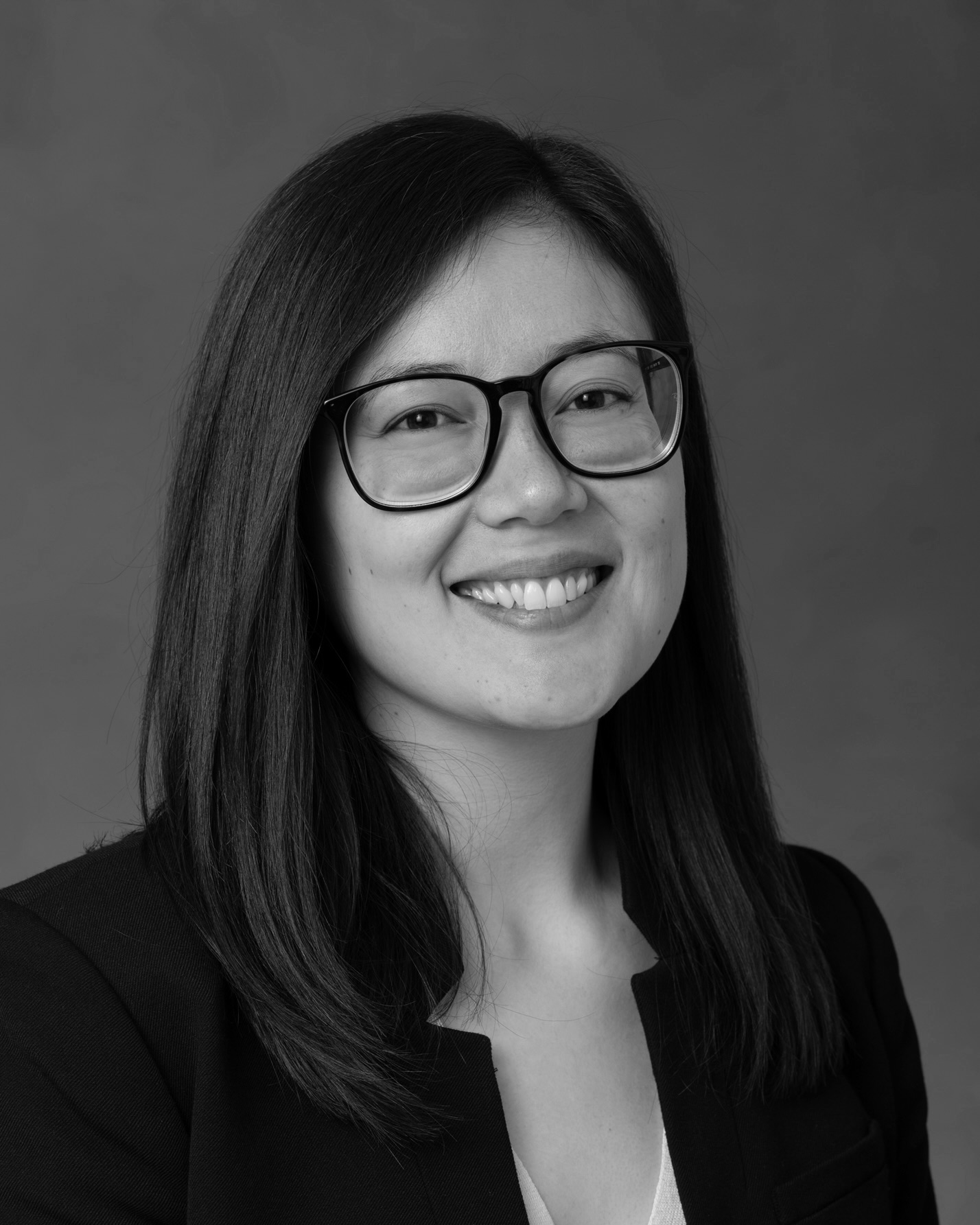}
received her B.S. and PhD degrees in Optical Science and Engineering from the University of Arizona, College of Optical Sciences. She is an Optical Engineer and Technical Group Supervisor at Jet Propulsion Laboratory in Pasadena, CA. She most recently served as the Optics Lead of the Earth Surface Mineral Dust Source Investigation (EMIT) instrument. She was awarded the JPL Charles Elachi Award and a NASA Honor Achievement Medal for Exceptional Engineering in the development of novel alignment methodologies for EMIT in 2023. 
\end{biographywithpic} 

\begin{biographywithpic}
{Christian Frankenberg}{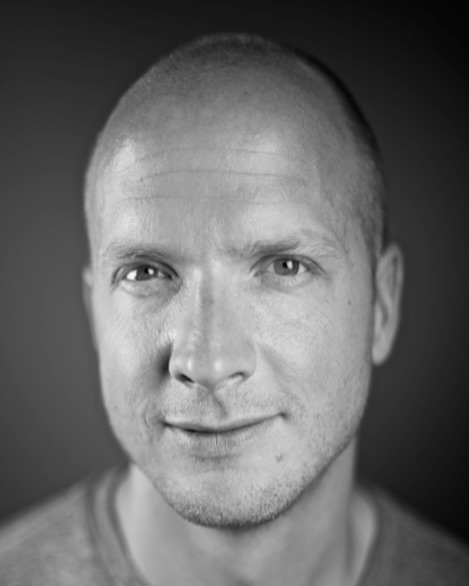}
received a Diploma from the University of Bayreuth and a PhD from the University of Heidelberg, Germany. He is a Professor of Environmental Science and Engineering at Caltech and Research Scientist at the Jet Propulsion Laboratory. His research is focused on studying the global carbon cycle from the vantage point in space and he has worked with various satellite missions and developed retrieval algorithms for greenhouse gas observations from space.
\end{biographywithpic}

\begin{biographywithpic}
{Rami Wehbe}{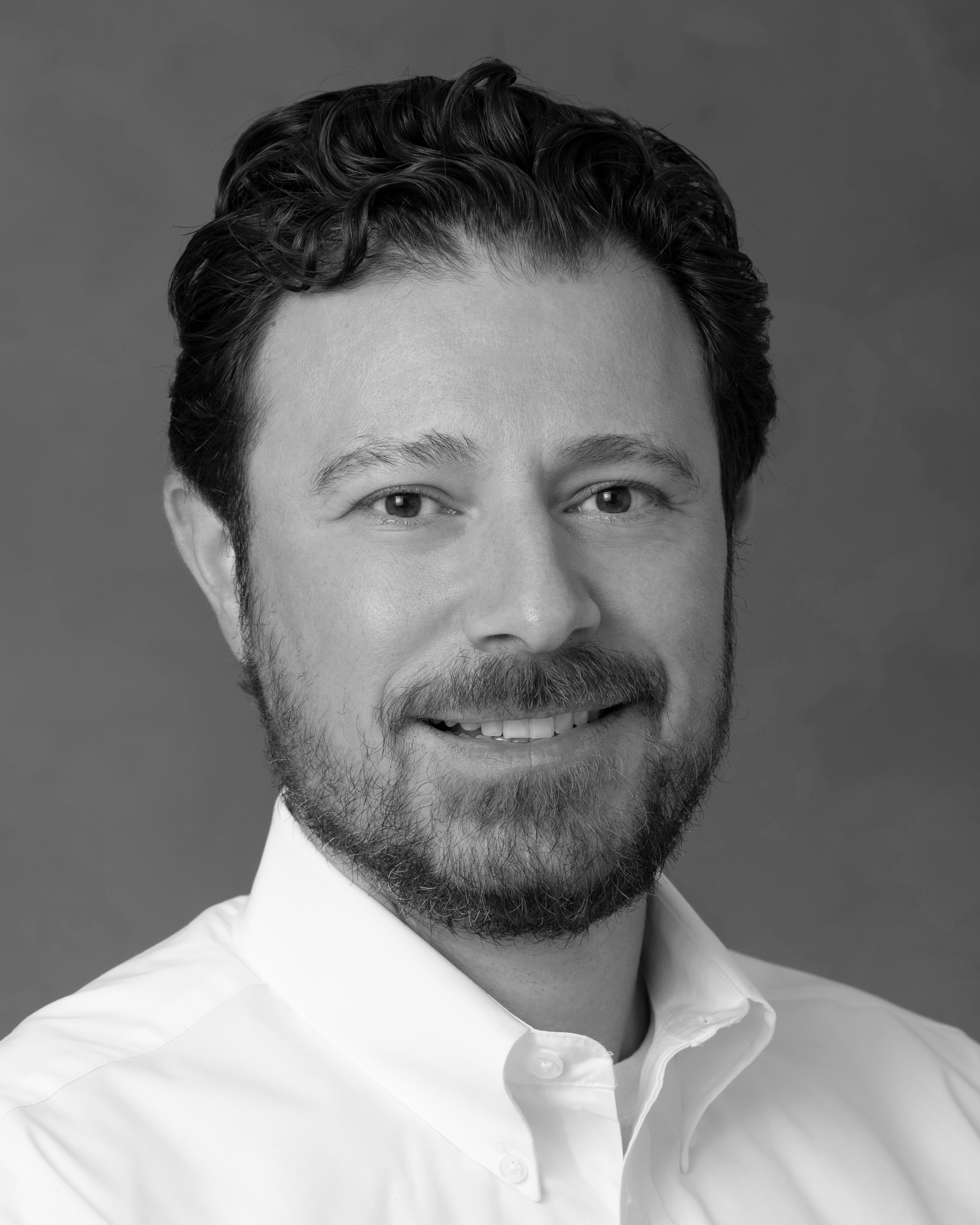}
received his B.S. in Mechanical Engineering from California State Polytechnic University, Pomona, a MS degree in Systems Engineering from Johns Hopkins University, and an MBA from the University of Southern California. He is an Optomechanical Engineer and Technical Group Supervisor at Jet Propulsion Laboratory in Pasadena, CA. He has designed and built airborne imaging spectrometers and contributed to SHERLOC and MISE flight missions.
\end{biographywithpic}

\begin{biographywithpic}
{Matthew Smith}{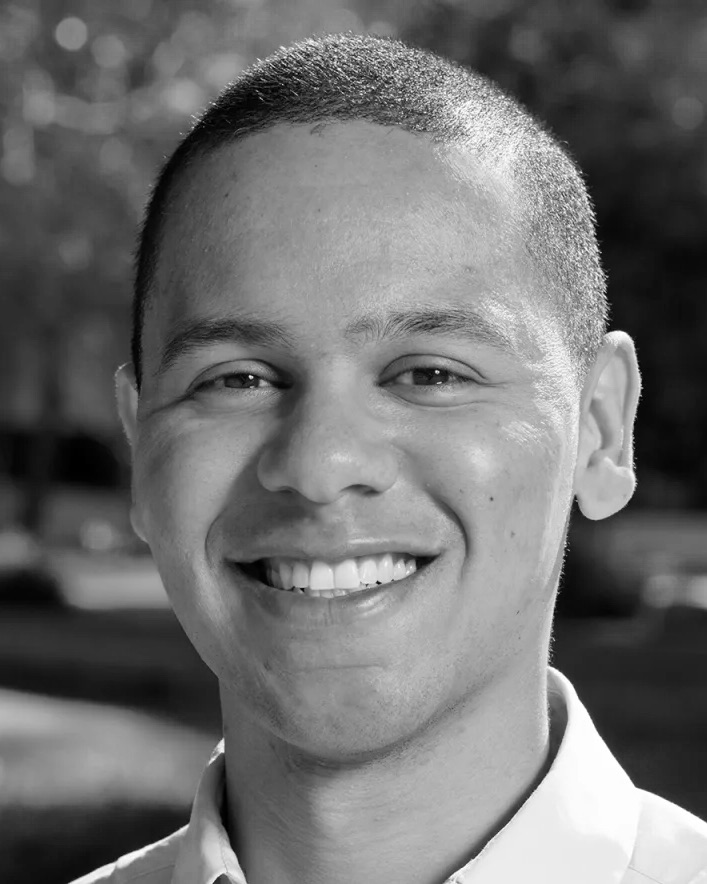}
received his BS degree in Applied Math from Harvard University and his MS and PhD degrees in Aeronautics and Astronautics from the Massachusetts Institute of Technology. He is an Instrument Systems Engineer at the Jet Propulsion Laboratory in Pasadena, CA.   
\end{biographywithpic}

\begin{biographywithpic}
{Sharmila Padmanabhan }{Bio_Sharmila.jpg}
(Senior Member, IEEE) received the Ph.D. degree from Colorado State University, Fort Collins, CO in 2009. She is a group supervisor of the instrument architecture and systems engineering group with the Instrument Electronics and Software Section at the Jet Propulsion Laboratory, California Institute of Technology, Pasadena, CA. Her research interests include millimeter, submillimeter-wave, and infrared instrumentation for remote sensing, calibration, validation, and performance assessment of microwave radiometers, and geophysical retrieval algorithm development. Dr. Padmanabhan was a recipient of the NASA Exceptional Public Achievement Medal in 2019.   
\end{biographywithpic}

\begin{biographywithpic}
{Valerie Scott}{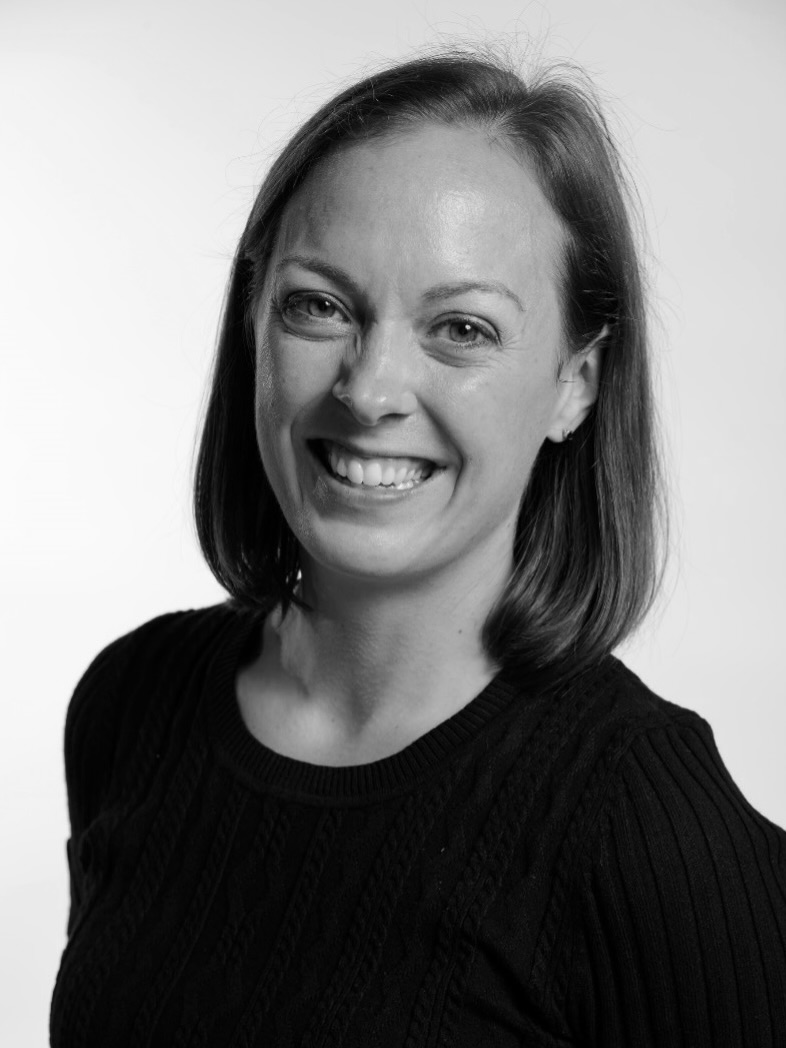}
is an engineer by profession and a scientist by training, with a career-long focus on bridging the gap between these two communities. Her experience spans technology development, from harsh environment components to sensors for a “tactile” wheel to full instruments such as a combination Mössbauer and X-ray fluorescence spectrometer and portfolio management, as Earth Science Technology Office Associate and Chevron Technical Fellow.  She currently has an emphasis on mission formulation, including as Capture Lead for Carbon-I.  
\end{biographywithpic}

\begin{biographywithpic}
{David R. Thompson}{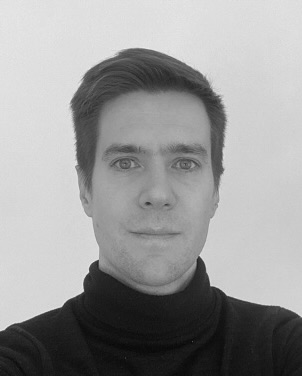}
is a Senior Research Scientist at the Jet Propulsion Laboratory, California Institute of Technology.  His research advances the algorithms and practice of imaging spectroscopy for characterizing Earth and other planetary bodies.  He is the Instrument Scientist for NASA’s EMIT and Lunar Trailblazer missions, and Deputy Project Scientist for NASA’s SBG-VSWIR imaging spectrometer mission.  He has received the NASA Exceptional Technology Achievement Medal, the NASA Exceptional Engineering Achievement Medal, the Lew Allen Award for Excellence, and the NASA Early Career Achievement Medal.  
\end{biographywithpic}

\begin{biographywithpic}
{Daniel W. Wilson}{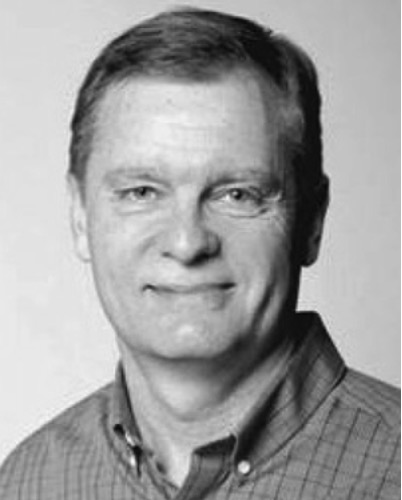}
received his Ph.D. in electrical engineering from the Georgia Institute of Technology in 2004. Since 1994, he has been working with the Microdevices laboratory, Jet Propulsion Laboratory (HPL) in Pasadena, CA, where he has focused on the design, modeling, and electron-beam fabrication of diffractive optical components and novel optical instruments. He is currently a Principal Engineer and Supervisor with the E-beam Technologies Group, JPL. He has been a key contributor to the successful development of tailored-efficiency curved diffraction gratings, coronagraph occulting masks, transient-event imaging spectrometers, and particle velocity sensors. He has delivered high-performance gratings for many spaceborne and airborne compact imaging spectrometers, Dr. Wilson has won the NASA Exceptional Technology Achievement Medal in 2010 for his work on electron-beam fabricated gratings.  
\end{biographywithpic}

\begin{biographywithpic}
{Pantazis Mouroulis}{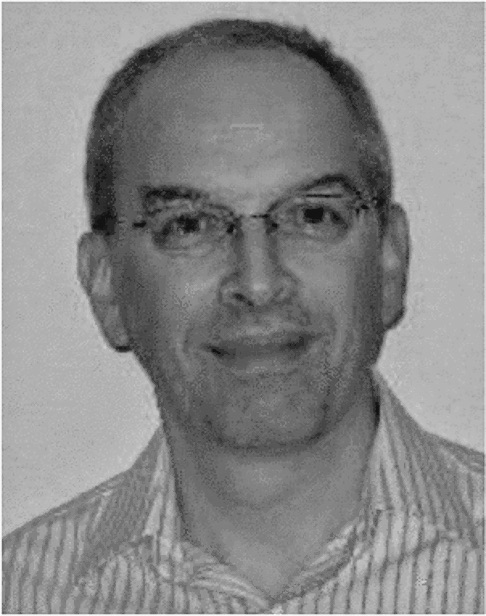}
is a Senior Research Scientist at Jet Propulsion laboratory with a 20-year focus on designing state-of-the-art imaging spectrometers for NASA and our nation. He is the Principal Investigator of the PRISM, SWIS, and VMDIS imaging spectrometers at JPL.
\end{biographywithpic}

\begin{biographywithpic}
{Robert O. Green}{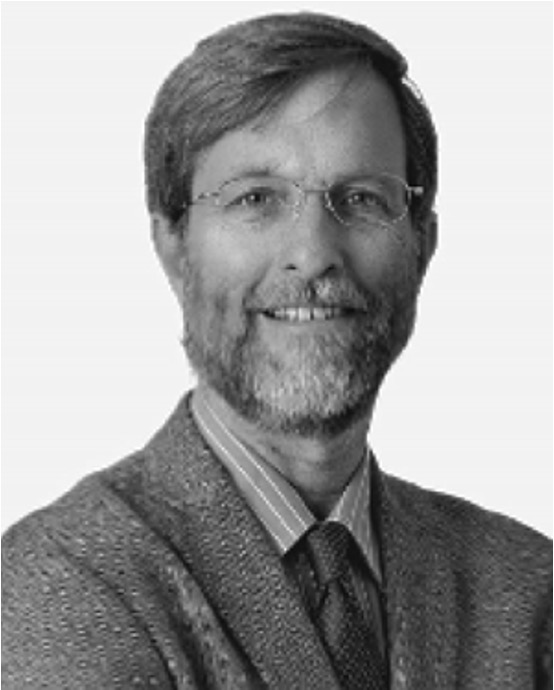}
is a Senior Research Scientist at Jet Propulsion Laboratory with a 30-year focus on using advanced imaging spectroscopy for science and discovery on Earth and through the solar system. Among other roles, he is the Principal Investigator of EMIT and the AVIRIS series of instruments.
\end{biographywithpic}

\end{document}